\documentclass[preprint]{aastex}
\usepackage{amssymb}
\usepackage{stmaryrd}
\usepackage{amssymb,amsmath, epsfig, epsf, lscape}

\def\beq{\begin{equation}}
\def\eeq{\end{equation}}
\def\bey{\begin{eqnarray}}
\def\eey{\end{eqnarray}}

\def\mpc{\, h^{-1}{\rm {Mpc}}}

\def\mpci{\, h{\rm {Mpc}}^{-1}}
\def\kpc{\, h^{-1}{\rm {kpc}}}

\def\msun{\, h^{-1}{\rm M_\odot}}

\def\g
 s{\mathrel{\raise1.16pt\hbox{$>$}\kern-7.0pt
\lower3.06pt\hbox{{$\scriptstyle \sim$}}}}
\def\ls{\mathrel{\raise1.16pt\hbox{$<$}\kern-7.0pt
\lower3.06pt\hbox{{$\scriptstyle \sim$}}}}
\def\gtsima{$\; \buildrel > \over \sim \;$}
\def\ltsima{$\; \buildrel < \over \sim \;$}
\def\prosima{$\; \buildrel \propto \over \sim \;$}
\def\gsim{\lower.5ex\hbox{\gtsima}}
\def\lsim{\lower.5ex\hbox{\ltsima}}
\def\simgt{\lower.5ex\hbox{\gtsima}}
\def\simlt{\lower.5ex\hbox{\ltsima}}
\def\simpr{\lower.5ex\hbox{\prosima}}
\def\la{\lsim}

\newcommand{\etal}{{et al.~}}

\slugcomment{To appear in {\it The Astrophysical Journal}}
\righthead{Reconstructing the Initial Density Field of the Local
Universe} \shorttitle{Reconstructing the Initial Density Field of
the Local Universe} \shortauthors{Wang H.Y. et al.}
\received{2013}

\begin{document}
\title {Reconstructing the Initial Density Field of the Local Universe: Method and Test with Mock Catalogs}
\author{Huiyuan Wang\altaffilmark{1,2}, H.J. Mo\altaffilmark{2}, Xiaohu Yang\altaffilmark{3,4} and Frank
C. van den Bosch\altaffilmark{5}}

\altaffiltext{1}{Key Laboratory for Research in Galaxies and
Cosmology, Department of Astronomy, University of Science and
Technology of China, Hefei, Anhui 230026, China;
whywang@mail.ustc.edu.cn} \altaffiltext{2}{Department of
Astronomy, University of Massachusetts, Amherst MA 01003-9305,
USA} \altaffiltext{3}{Key Laboratory for Research in Galaxies and
Cosmology, Shanghai Astronomical Observatory, Shanghai 200030,
China} \altaffiltext{4}{Center for Astronomy and Astrophysics,
Shanghai Jiao Tong University, Shanghai 200240, China}
\altaffiltext{5}{Astronomy Department, Yale University, P.O. Box
208101, New Haven, CT 06520-8101, USA}

\begin{abstract}
Our research objective in this paper is to reconstruct an initial
linear density field, which follows the multivariate Gaussian
distribution with variances given by the linear power spectrum of
the current CDM model and evolves through gravitational
instability to the present-day density field in the local
Universe. For this purpose, we develop a Hamiltonian Markov Chain
Monte Carlo method to obtain the linear density field from a
posterior probability function that consists of two components: a
prior of a Gaussian density field with a given linear spectrum,
and a likelihood term that is given by the current density field.
The present-day density field can be reconstructed from galaxy
groups using the method developed in Wang et al. (2009a). Using a
realistic mock SDSS DR7, obtained by populating dark matter haloes
in the Millennium simulation with galaxies, we show that our
method can effectively and accurately recover both the amplitudes
and phases of the initial, linear density field. To
examine the accuracy of our method, we use $N$-body simulations
to evolve these reconstructed initial conditions to the present
day. The resimulated density field thus obtained accurately
matches the original density field of the Millennium simulation in
the density range $0.3 \la \rho/\bar{\rho} \la 20$ without any
significant bias. Especially, the Fourier phases of the
resimulated density fields are tightly correlated with those of
the original simulation down to a scale corresponding to a
wavenumber of $\sim 1\mpci$, much smaller than the translinear
scale, which corresponds to a wavenumber of $\sim 0.15\mpci$.
\end{abstract}

\keywords{dark matter - large-scale structure of the universe -
galaxies: haloes - methods: statistical}

\section{Introduction}
\label{sec_intro}

In the current cold dark matter (CDM) cosmogony, a key concept in
the build-up of structure is the formation of dark matter haloes.
These are quasi-equilibrium systems of dark matter, formed through
non-linear gravitational collapse. In a CDM-like hierarchical
scenario, most of the mass is bound within haloes; galaxies and
other luminous objects are assumed to form in these haloes because
of cooling and condensation of baryonic gas (see Mo, van den Bosch
\& White 2010). With $N$-body simulations, the properties of the
halo population, such as the spatial clustering properties, the
mass function, the assembly histories and the internal structures
are well understood. Nevertheless, how galaxies form in dark
matter haloes in the cosmic density field remains an unsolved
problem. A long-standing problem in current galaxy formation
theory is to explain the low efficiency with which baryonic gas is
converted into stars: the observed mass in stars at the present
time is less than $10\%$ of the total baryonic mass in the
universe (Bell et al. 2003). Including cold gas associated with
galaxies only increases this to $\sim 12\%$. This low efficiency
of star formation and gas assembly into galaxies is not a natural
consequence of hierarchical formation, in which the gas is
expected to cool rapidly at high redshift in low-mass dark matter
haloes. A number of physical processes have been proposed to
suppress gas cooling and the star formation efficiency. These
include photoionization heating by the UV background (e.g.
Efstathiou 1992; Thoul \& Weinberg 1996; Somerville 2002; Gnedin
2000; Hoeft et al. 2006), feedback from supernova explosions (e.g.
White \& Rees 1978; Dekel \& Silk 1986) and from AGN (e.g. Tabor
\& Binney 1993; Ciotti \& Ostriker 1997; Hopkins et al. 2006 and
references therein; Wang et al. 2012b), and pre-heating by star
formation/AGN (e.g. Mo \& Mao 2002; Pfroemmer et al. 2012), and by
pre-virialization (Mo et al. 2005). Unfortunately, our
understanding of all these processes is still poor, making it
difficult to test the predictions of these scenarios with
observations.

In order to understand the galaxy formation processes
throughout the cosmic density field, a key step is to study the
distributions and properties of galaxies and the intergalactic
medium (IGM), and their interactions with each other and with dark
matter. In the local universe, detailed observations of the galaxy
population are now available from large redshift surveys, for
example the Sloan Digital Sky Survey (SDSS; York et al. 2000).
This not only allows us to study the large-scale structure in the
local universe, but can also be used to derive a large number of
physical quantities characterizing the intrinsic properties of
individual galaxies, such as luminosity, stellar mass, color,
morphology, size, star formation rate, and nuclear activity. There
have also been observational programs dedicated to the various
aspects of the IGM. Extensive X-ray observations have been
conducted to study the hot gas associated with clusters and rich
groups of galaxies but the total gas mass associated with
these systems is expected to be small. With the advent of
accurate measurements of the cosmic microwave background from
observations such as the South Pole Telescope, the Atacama
Cosmology Telescope, and the PLANCK Satellite, one can also probe
the hot, diffuse gas outside clusters and groups through the
Sunyaev-Zel'dovich (SZ) effect. However, at low redshift,
about $70\%$ of all the mass is in virialized halos with virial
temperatures below $10^{6}$ K (see Mo \& White 2002), too cold to
be studied with X-ray data and/or the SZ effect. A promising way
to study the diffuse IGM at such low temperature is through quasar
absorption lines. With the installation of the {\it Cosmic Origins
Spectrograph} on the {\it HST} the sample of UV absorption systems
at low redshift is expected to increase by an order-of-magnitude
or more, allowing a much more detailed examination of the warm
component of the local IGM.

These observational programs together provide an unprecedented data
base to study how galaxies form and evolve in the cosmic density
field. However, in order to make full use of the potential of the
observational data to test models, one has to develop an optimal
strategy. Conventionally, one starts with a cosmological model, {\it
  e.g.} the current $\Lambda$CDM model, uses computer simulations to
follow the evolution of the cosmic density field, and compares
simulation results with observational data in a statistical way.
However, such a comparison can be made directly only under the
assumption that the observational sample and simulation are fair
representations of the universe, so that cosmic variance is not an
issue. Unfortunately, this assumption is almost always violated in
reality. Simulations are limited by the dynamical ranges they can
cover. In order to resolve processes on the scale of galaxies, the
simulation volume has often to be much smaller than a fair sample
of the large scale structure. Observationally, finite sample
volumes also lead to a biased representation of the statistical
properties of the cosmic density field and galaxy population in
the universe.

It is thus imperative to have theoretical and
empirical input to optimize an observational strategy and to help
interpret the limited observations in an unbiased way. The
uncertainties can be minimized if comparisons between observation
and model prediction are made for systems that have both the same
environments and the same formation histories. Ideally, if we can
accurately reconstruct the initial conditions for the formation of
the structures in which the observed galaxy population resides and
from which the actual gas emissions and absorptions are produced,
then we will be able to compare observation and simulation (i.e.,
data and theory) in a way that is free of cosmic variance, thereby
greatly enhancing the constraining power of the observational
data.

The goal of this paper is to develop a method that can be used to
reconstruct the initial (linear) density field that forms the
large scale structure in the local universe. In this first paper
in a series, we describe our reconstruction method and test its
performance with realistic mock galaxy catalogs. The structure of
the paper is arranged as follows. In Section~\ref{sec:method}, we
describe our reconstruction method. In Section~\ref{sec:sim}, we
test our method using a simulated density field. In
Section~\ref{sec:mocktest} we present our mock catalog that is
used to test our method and our results of the linear density
field reconstructed from it. In Section \ref{sec:resim} we use
N-body simulations to follow the structure formation seeded by the
re-constructed initial density field, and compare the final
density field with the original one used to construct the mock
catalogs. Finally Section \ref{sec:discussion} contains a summary
of the main results and some further discussions.

In order to avoid confusion, we here list the various matter
density fields used in the text: (i) The final density field,
$\rho_{\rm f}({\bf x})$: the \emph{true} present-day (final)
density field, either in the real Universe or in an original
$N$-body simulation. (ii) The reconstructed density field,
$\rho_{\rm rc}({\bf x})$: the present-day density field
re-constructed from the mock catalog. (iii) The reconstructed
initial (linear) density field: the initial linear density
re-constructed from a present-day density field.  (iv) The
re-simulated density field, $\rho_{\rm rs}({\bf x})$: the
present-day density field obtained from numerical $N$-body
simulations using the re-constructed initial density field as
initial conditions. (v) The modeled density field, $\rho_{\rm
mod}({\bf x})$: a model prediction for the present-day density
field obtained from the initial density field using the modified
Zel`dovich approximation introduced by Tassev \& Zaldarriaga
(2012b, hereafter TZ12).  This model density field is used in the
reconstruction method to link the initial and final density
fields.

\section{Method}
\label{sec:method}

Our reconstruction consists of the following several steps: (i)
Use galaxy groups\footnote{We use galaxy group to refer a galaxy
system (a cluster or a group) without regard to its richness.}
selected from the SDSS, to represent dark matter haloes; (ii) Use
haloes above a certain mass to reconstruct the cosmic density
field at the present day; (iii) Reconstruct the initial density
field that best matches the final density field under the
constraint of current cosmology and a linear perturbation
spectrum. The galaxy group finder used is described and tested in
detail in Yang et al. (2005, 2007). The method for reconstructing
the density field starting from dark matter haloes (i.e., galaxy
groups) above a given mass threshold is described and tested in
Wang et al. (2009a, 2012a). In what follows we describe how we
reconstruct the initial linear density field from a given
present-day density field.

\subsection{Objectives and the Posterior Probability
Distribution}\label{sec_goal}

Our goal is to obtain the linear density field that can reproduce
a given present-day density field. We work in Fourier space, so
that the initial density field is specified by $\delta({\bf k})$,
the Fourier modes of the initial density field. Two constraints
are used in the reconstruction. First, according to the standard
cosmology, we assume the linear density field to be Gaussian, so
that the Fourier modes obey the following probability
distribution:
\begin{equation}\label{eq_gs}
P[\delta_j({\bf k})]=\prod_{\bf k}^{\rm half} \prod_{j=0}^1
\frac{1}{[\pi P_{\rm lin}(k)]^{1/2}} \exp\left\{-[\delta_j({\bf
k})]^2/P_{\rm lin}(k)\right\}\,,
\end{equation}
where $P_{\rm lin}(k)$ is the (analytical) linear power spectrum,
and the subscripts $j = 0,1$ denote the real and imaginary parts,
respectively. Because $\delta({\bf k})$ is the Fourier transform
of a real field, $\delta({\bf k}) = \delta^{\ast}(-{\bf k})$ so
that only the Fourier modes in the upper half-space (i.e. with
$k_z\geq 0$) are needed.  Second, the density field, $\rho_{\rm
mod}({\bf x})$, evolved from this linear density field according
to a chosen model of structure formation, should best match a
present-day density field, $\rho_{\rm p}({\bf
  x})$. In other words, we seek the appropriate $\delta({\bf k})$ to
minimize a `cost parameter' which we define as
\begin{equation}
\chi^2=\sum_{\bf x}\frac{[\rho_{\rm mod}({\bf x})-\rho_{\rm
p}({\bf x})]^2\omega({\bf x})}{2\sigma_{\rm p}^2({\bf x})}\,,
\end{equation}
where $\sigma_{\rm p}({\bf x})$ is the statistical uncertainties
in $\rho_{\rm p}({\bf x})$, while $\omega({\bf x})$ is a weight
function used to account for the survey geometry. The
present-day density field, $\rho_{\rm p}$, may either be the
original simulated density field ($\rho_{\rm f}$, Section
\ref{sec:sim}) or the density field reconstructed from a galaxy
redshift survey ($\rho_{\rm rc}$, Section \ref{sec:mocktest}).
The uncertainties $\sigma_{\rm p}({\bf x})$ are found to be
roughly proportional to $\rho_{\rm p}({\bf x})$ (see Section
\ref{sec_denrc}), and so we set $\sigma_{\rm p}({\bf x}) =
\mu\rho_{\rm p}({\bf x})$, with $\mu$ a constant parameter. In
order to obtain the model prediction for the final density field,
we need a model to link $\rho_{\rm mod}({\bf x})$ with
$\delta({\bf k})$. This model should not only be accurate, but also be efficient
so that the computation can be achieved in a reasonable amount of
time, as to be discussed in Section \ref{sec_hf}. In practice, all
these fields are to be sampled in a periodic box of length $L$ on a
side, divided into $N_{\rm c}$ grids in each dimension, so that the
number of Fourier modes to be dealt with is finite.

Because of the statistical uncertainties in $\rho_{\rm p}({\bf
x})$ and the finite survey volume, the solution for $\delta({\bf
k})$ under the two constraints described above is not unique, but
should obey the posterior probability distribution of $\delta({\bf
k})$. Assuming that the likelihood of $\rho_{\rm mod}({\bf x})$
given $\rho_{\rm p}({\bf x})$ is $\exp(-\chi^2)$, the posterior
probability distribution for $\delta({\bf k})$ given $\rho_{\rm
p}({\bf x})$ can be written as
\begin{equation}
\emph{Q}(\delta_j({\bf k})|\rho_{\rm p}({\bf x}))={\rm
e}^{-\sum_{\bf x}[\rho_{\rm mod}({\bf x})-\rho_{\rm p}({\bf
x})]^2\omega({\bf x})/2\sigma_{\rm p}^2({\bf x})}\prod_{\bf
k}^{\rm half}\prod_{j=0}^1\frac{1}{[\pi P_{\rm lin}(k)]^{1/2}}{\rm
e}^{-[\delta_j({\bf k})]^2/P_{\rm lin}(k)}\label{eq_post}\,.
\end{equation}
For our purpose, we seek the solutions for $\delta({\bf k})$
that maximize this posterior probability distribution function.

We use the Hamiltonian Markov Chain Monte Carlo technique (HMC) to
achieve our goal. The HMC method was originally developed to
sample a posterior distribution (Duane et al. 1987; Neal 1996) and
has proven to be effective for exploring large, multi-dimensional
posterior spaces (e.g. Hanson 2001). Different from the
conventional Markov Chain Monte Carlo, the HMC method introduces a
persistent motion of the Markov Chain when exploring the parameter
space so that the random walk is greatly suppressed and the
efficiency much improved (Duane et al. 1987). This method
has already been widely used in astrophysics and cosmology (see
Hajian 2007; Taylor, Ashdown \& Hobson 2008; Jasche \& Kitaura
2010; Kitaura et al. 2012a; Jasche \& Wandelt 2013, hereafter
JW13; Kitaura 2013). For example, both JW13 and Kitaura (2013)
developed methods incorporating the HMC to reconstruct the initial density field from galaxy distribution. In particular, JW13 used a
posterior distribution function composed of a Poissonian likelihood
based directly on galaxy distribution and a prior distribution of the initial
density field. They successfully drew a sample of the initial
density field from their posterior distribution, demonstrating
that HMC is a powerful method for reconstructing the initial
density field.  In principle, such a sample can be used to inspect
the statistical uncertainties in the reconstruction. However, this
kind of analysis requires a careful design of the likelihood
function to take into account in detail the statistical
uncertainties in the constraining data and in the model of the
cosmic density field. Our basic idea is similar to that of JW13,
but for our purpose described above we restrict ourselves to
seeking the maximum posterior estimates of $\delta({\bf k})$
instead of obtaining the posterior distribution of $\delta({\bf
k})$. As we will demonstrate below using realistic mock catalogs
constructed from a cosmological $N$-body simulation, our HMC
method based on the likelihood function defined above is
sufficient for this research objective.

\subsection{The Hamiltonian Monte Carlo Method}

In this subsection, we briefly outline the HMC method (see Hanson
2001; Taylor et al. 2008; JW13 for some more detailed
descriptions). The method is itself based on an analogy to solving
a physical system in Hamiltonian dynamics. As a first step, we
define the potential of the system to be the negative of the
logarithm of the target probability distribution,
\begin{eqnarray}\label{targetprob}
\psi[\delta_j({\bf k})]& \equiv &-\ln[\emph{Q}(\delta_j({\bf
k})|\rho_{\rm p}({\bf x}))]\nonumber\\
&=&\sum_{\bf k}^{\rm half}\ln[\pi P_{\rm lin}(k)]+\sum_{\bf
k}^{\rm half}\sum_{j=0}^1\frac{[\delta_j({\bf k})]^2}{P_{\rm
lin}(k)}+\sum_{\bf x}\frac{[\rho_{\rm mod}({\bf x})-\rho_{\rm
p}({\bf x})]^2\omega({\bf x})}{2\sigma_{\rm p}^2({\bf x})}\,.
\end{eqnarray}
For each $\delta_j({\bf k})$, a momentum variable, $p_j({\bf k})$, and
a mass variable, $m_j({\bf k})$, are introduced. The Hamiltonian of
the fictitious system can then be written as
\begin{equation}
H=\sum_{\bf k}^{\rm half}\sum_{j=0}^1\frac{p_j^2({\bf
k})}{2m_j({\bf k})}+\psi[\delta_j({\bf k})]\,.
\end{equation}
The statistical properties of the system is given by the partition
function, $\exp(-H)$, which can be separated into a Gaussian
distribution in momenta ${p_j({\bf k})}$ multiplied by the target
distribution,
\begin{equation}
\exp(-H)=\emph{Q}[\delta_j({\bf k})|\rho_{\rm p}({\bf x})]
\prod_{\bf k}^{\rm half}\prod_{j=0}^1{\rm e}^{-\frac{p_j^2({\bf
k})}{2m_j({\bf k})}}\label{eq_eh}\,.
\end{equation}
Thus, the target probability distribution can be obtained by first
sampling this partition function and then marginalizing over
momenta (i.e setting all the momenta to be zero).

In order to sample from the partition function, we first pick
a set of momenta $p_j({\bf k})$ randomly from the
multi-dimensional un-correlated Gaussian distribution with
variances $m_j({\bf k})$. We describe how to pick the mass
variables in Section \ref{sec_Hmass}. We then evolve the system
from the starting point $[\delta_j({\bf k}), p_j({\bf k})]$ in the
phase space to some pseudo time $T$ according to the Hamilton
equations,
\begin{equation}
\frac{{\rm d}\delta_j({\bf k})}{{\rm d}t}=\frac{\partial
H}{\partial p_j(\bf k)}=\frac{p_j(\bf k)}{m_j(\bf k)}\,;
\end{equation}
\begin{equation}
\frac{{\rm d}p_j({\bf k})}{{\rm d}t}=-\frac{\partial H}{\partial
\delta_j(\bf k)}=-\frac{2\delta_j({\bf k})}{P_{\rm lin}(k)}-F_{j}({\bf k})\,,
\label{eq_dp}
\end{equation}
where $F_j({\bf k}) = \partial\chi^2/\partial\delta_j({\bf k})$ is
the likelihood term of the Hamiltonian force to be discussed
below. The integrated trajectory finally reaches a point
$[\delta'_j({\bf k}),p'_j({\bf k})]$ in phase space and we accept
this state with a probability
\begin{equation}\label{paccept}
p={\rm min}\left \{1, {\rm e}^{-[H(\delta'_j({\bf k}), p'_j({\bf
k})) -H(\delta_j({\bf k}), p_j({\bf k}))]}\right \}\,.
\end{equation}
The procedure is repeated by randomly picking a new set of
momenta.

Since the Hamiltonian of a physical system is conserved, the
acceptance rate should in principle be unity, which is one of the
main advantages of working with  the partition function,
$\exp(-H)$, instead of the target distribution function itself.
However, rejection may occur because of numerical errors. In order
to optimize the acceptance rate, it is common practice to
integrate the ``equations of motion'' using the leapfrog
technique,
\begin{equation}\label{eq_lf1}
p_j({\bf k},t+\tau/2)=p_j({\bf k},t)-\frac{\tau}{2}\frac{\partial
H}{\partial \delta_j(\bf k)}{\Bigg\vert}_t\,;
\end{equation}
\begin{equation}\label{eq_lf2}
\delta_j({\bf k},t+\tau)=\delta_j({\bf k},t)+\frac{\tau}{m_j({\bf
k})}p_j({\bf k},t+\tau/2)\,;
\end{equation}
\begin{equation}\label{eq_lf3}
p_j({\bf k},t+\tau)=p_j({\bf
k},t+\tau/2)-\frac{\tau}{2}\frac{\partial H}{\partial
\delta_j(\bf k)}{\Bigg\vert}_{t+\tau}\,,
\end{equation}
where $\tau$ represents the time increment for the leapfrog step.
The leapfrog technique uses half-time steps in the first and third
equations so that the scheme is accurate to second order in
$\tau$. The equations are integrated for $n$ steps so that
$n\tau=T$. The value of $T$ must be randomized to avoid resonance
trajectories. We thus randomly pick $n$ and $\tau$ from two
uniform distributions in the range of $[1,n_{\rm max}]$ and
$[0,\tau_{\rm
  max}]$, respectively. We will discuss our choices of $n_{\rm max}$
and $\tau_{\rm max}$ below. The $n$ leapfrog steps are referred to
as one chain step.

\subsection{Hamiltonian Force and Structure Formation Model}
\label{sec_hf}

As shown in Eq. (\ref{eq_dp}), the Hamiltonian force consists of two
components, the prior term, $2\delta_j({\bf k})/P_{\rm lin}(k)$, and the
likelihood term, $F_{j}({\bf k})$. The latter can be re-written as:
\begin{eqnarray}\label{eq_pi1}
F_{j}({\bf k})&=&\sum_{\bf x}\frac{[\rho_{\rm mod}({\bf
x})-\rho_{\rm p}({\bf x})]\omega({\bf x})}{\sigma_{\rm p}^2({\bf
x})}\frac{\partial\rho_{\rm mod}({\bf x})}{\partial\delta_j({\bf
k})}\equiv\sum_{\bf x}\rho_{\rm d}({\bf
x})\frac{\partial\rho_{\rm mod}({\bf x})}{\partial\delta_j({\bf k})}\nonumber\\
&=&\sum_{\bf x}\sum_{\bf k_2}\rho_{\rm d}({\bf k_2}){\rm e}^{i{\bf
k_2\cdot x}}\frac{\partial}{\partial\delta_j({\bf k})}\sum_{\bf
k_1}\rho_{\rm mod}({\bf k_1}){\rm e}^{i{\bf k_1\cdot x}}=N_{\rm c}^3\sum_{\bf
k_1}\rho_{\rm d}^{\ast}({\bf k_1})\frac{\partial\rho_{\rm mod}({\bf
k_1})}{\partial\delta_j({\bf k})}\,.
\end{eqnarray}
For the sake of simplicity we have introduced in this equation a
new quantity, $\rho_{\rm d}({\bf x})$, which is directly related to
$\rho_{\rm mod}({\bf x})$, as defined in the second equation.

To derive the solution of the Hamiltonian force, we need a model
of structure formation to connect $\rho_{\rm mod}({\bf k})$ and
$\delta({\bf k})$. In this work, we adopt the model developed in
TZ12. According to TZ12, the present-day density field can be
written in terms of the modeled density field, $\rho_{\rm mod}$,
as $\rho_{\rm p}({\bf k}) = \rho_{\rm mod}({\bf k}) + \rho_{\rm
mc}({\bf k})$. Here, $\rho_{\rm mc}({\bf k})$, a random
mode-coupling residual, is generally small on mildly non-linear
scales and can be neglected. The modeled density can be obtained
via $\rho_{\rm mod}({\bf k}) = R_{\delta}(k)\rho_{\rm MZ}({\bf
k})$, where $\rho_{\rm MZ}$ is the density field predicted by the
Modified Zel'dovich approximation (MZA) developed by TZ12 and
$R_{\delta}(k)$ is a density transfer function. The transfer
function is obtained via comparing the prediction of the
Zel'dovich approximation with $N$-body simulations, and
can suppress the effects of shell crossing. Using numerical
simulations as reference, TZ12 found that the prediction of MZA is
better/worse than that of the second-order Lagrangian perturbation
theory (2LPT) on small/large scales. Because MZA is computationally faster than 2LPT, we choose MZA to predict the present-day density field.
Following TZ12, we use the MZA to derive the displacement field,
${\bf s}({\bf q})$,
\begin{equation}\label{eq_disp}
{\bf s}({\bf q})=\sum_{\bf k_2}{\bf s(k_2)}{\rm e}^{i{\bf k_2\cdot
q}}=\sum_{\bf k_2}R_z(k_2){\bf s_z(k_2)}{\rm e}^{i{\bf k_2\cdot
q}}=\sum_{\bf k_2}R_z(k_2)\frac{i\bf k_2}{k^2_2}\delta({\bf
k_2}){\rm e}^{i{\bf k_2\cdot q}}\,,
\end{equation}
where ${\bf q}$ is a Lagrangian coordinate, ${\bf s_z(k_2)}$ is
the Zel'dovich displacement field, and
$R_z(k_2)=\exp(-0.085(k_2/k_{\rm NL})^2)$ is the transfer function
for the Zel'dovich displacement field, with $k_{\rm NL}$ the
non-linear scale at redshift zero. We move particles, which are
initially located on uniform grids of positions ${\bf q}$, to
${\bf x(q)}={\bf q}+{\bf s}({\bf q})$ to sample the density field.
We then utilize a clouds-in-cells (CIC) assignment (Hockney \&
Eastwood 1981) to construct the MZA density field on grids from
the particle population. We Fourier transform the MZA density, and
multiply it with the density transfer function
$R_{\delta}(k)=\exp(0.58d)$, where $d\equiv\delta^2(k/2)$,
\footnote{Both $k_{NL}$ and $\delta^2$ can be read off from Eq.
(3.9) in TZ12. The detailed form of the density transfer function
for MZA was communicated to us by Svetlin Tassev.} and a Gaussian
kernel $w_{\rm G}( R_{\rm s}k)$ characterized by a smoothing scale
$R_{\rm s}$. We then deconvolve the CIC kernel by dividing the
resulting density field in Fourier space by the Fourier transform
of the CIC kernel, $w_{\rm CIC}({\bf k})={\rm sinc}(k_xL/2N_{\rm
c}){\rm sinc}(k_yL/2N_{\rm c}){\rm sinc}(k_zL/2N_{\rm c})$, where
$k_x$, $k_y$ and $k_z$ are the $x$, $y$ and $z$ components of the
wavevector ${\bf k}$, respectively. Finally we obtain the modeled
density field as,
\begin{equation}\label{eq_rhoz}
\rho_{\rm mod}({\bf k_1})=\frac{R_{\delta}(k_1)w_{\rm G}(R_{\rm
s}k_1)}{N_{\rm c}^3}\sum_{\bf q}{\rm e}^{-i{\bf k_1\cdot x(q)}}\,.
\end{equation}
Inserting Eqs. (\ref{eq_disp}) and (\ref{eq_rhoz}) together with
${\bf x(q)}={\bf q}+{\bf s}({\bf q})$ into Eq. (\ref{eq_pi1}) then
yields the likelihood term of the Hamiltonian force:
\begin{eqnarray}\label{eq_pi2}
F_{j}({\bf k})&=&\sum_{\bf k_1}\rho_{\rm d}^{\ast}({\bf
k_1})R_{\delta}(k_1)w_{\rm G}(R_{\rm s}k_1)\sum_{\bf q}{\rm
e}^{-i{\bf k_1\cdot x}({\bf q})}\frac{\partial{(-i\bf k_1\cdot
x(q)})}{\partial\delta_j({\bf k})}\nonumber\\
&=&\sum_{\bf k_2}\frac{R_z(k_2)}{k_2^2} \frac{\partial\delta({\bf
k_2})}{\partial\delta_j({\bf k})}(i{\bf k_2\cdot})\sum_{\bf q}{\rm
e}^{i{\bf k_2\cdot q}}\sum_{\bf k_1}(-i{\bf k_1})\rho_{\rm
d}^{\ast}({\bf k_1})R_{\delta}(k_1)w_{\rm G}(R_{\rm s}k_1){\rm
e}^{-i{\bf k_1\cdot x(q)}}.
\end{eqnarray}
Note that $\partial{\bf x}/\partial\delta_j({\bf k})=\partial{\bf
s}/\partial\delta_j({\bf k})$ is used in the derivation.

For convenience we introduce a density-vector field,
\begin{equation}\label{eq_dvf}
{\bf \Psi}({\bf q}) \equiv N_{\rm c}^3\sum_{\bf k_1}(-i{\bf
k_1})\rho_{\rm d}^{\ast}({\bf k_1})R_{\delta}(k_1)w_{\rm G}(R_{\rm
s}k_1){\rm e}^{-i{\bf k_1\cdot x(q)}}.
\end{equation}
It is important to note that ${\bf \Psi}({\bf q})$ cannot be
directly derived with Fourier transformation because ${\bf x(q)}$
are not spaced regularly. To bypass this problem, we introduce a
transitional field in Fourier space, ${\bf \Gamma}({\bf
k_1})=N_{\rm c}^3(-i{\bf k_1})\rho_{\rm d}^{\ast}({\bf
k_1})R_{\delta}(k_1)w_{\rm G}(R_{\rm s}k_1)$, which is related to
the density-vector through ${\bf \Psi}({\bf q})={\bf
\Gamma}[{\bf-x(q)}]={\bf \Gamma}[{\bf L-x(q)}]$, with
${\bf\Gamma}({\bf x})$ the Fourier transform of ${\bf \Gamma}({\bf
k_1})$ and the vector ${\bf L}=(L,L,L)$. One might think that the
density-vector field can be derived straightforwardly via
interpolation. Unfortunately, interpolation can cause smoothing
and serious errors in the final estimation of the Hamiltonian
force. In order to correct these effects, we proceed as follows.
We first divide ${\bf \Gamma}({\bf k_1})$ by $w_{\rm CIC}({\bf
k_1})$ to deconvolve the CIC interpolation that is applied later.
We then Fourier transform the deconvolved ${\bf \Gamma}({\bf
k_1})$ into real space to obtain ${\bf \Gamma}'({\bf x})$. Finally
we interpolate ${\bf \Gamma}'({\bf x})$ to the position, ${\bf
L-x(q)}$, to obtain ${\bf \Psi}({\bf q})$ via a CIC scheme. We
emphasize again that deconvolving the CIC kernel is crucial for
obtaining an accurate estimate of the Hamiltonian force (see
Section \ref{sec:sim}). With ${\bf \Psi}({\bf q})$ obtained, we
can rewrite the Hamiltonian force as:
\begin{equation}\label{eq_hft}
F_{j}({\bf k})=\sum_{\bf
k_2}\frac{R_z(k_2)}{k_2^2}\frac{\partial\delta({\bf
k_2})}{\partial\delta_j({\bf k})}(i{\bf k_2\cdot})\sum_{\bf
q}\frac{1}{N_{\rm c}^3}{\rm e}^{i{\bf k_2\cdot q}}{\bf \Psi}({\bf
q})=\sum_{\bf k_2}\frac{R_z(k_2)}{k_2^2} \frac{\partial\delta({\bf
k_2})}{\partial\delta_j({\bf k})}(i{\bf k_2\cdot}){\bf
\Psi}^\ast({\bf k_2})\,,
\end{equation}
where ${\bf \Psi}({\bf k_2})$ is the Fourier transform of ${\bf
\Psi}({\bf q})$. Again, the last equation cannot be obtained
directly from Fourier transforms since the signs in the exponents
are not negative; it is obtained using the fact that ${\rm
e}^{i{\bf k_2\cdot q}}={\rm e}^{-i{\bf (-k_2)\cdot q}}$ and that
${\bf \Psi}(-{\bf k_2}) = {\bf\Psi}^\ast({\bf k_2})$.

Since $\partial\delta({\bf k_2})/\partial\delta_j({\bf k})$ is
nonzero only when ${\bf k_2=\pm k}$, the likelihood term of the
Hamiltonian force for the real part of $\delta({\bf k})$ can be
obtained as
\begin{equation}\label{eq_f0}
F_{0}({\bf k})=\frac{2R_z(k)}{k^2}{\bf k \cdot \Psi_1(k)}\,,
\end{equation}
and for the imaginary part as
\begin{equation}\label{eq_f1}
F_{1}({\bf k})=-\frac{2R_z(k)}{k^2}{\bf k\cdot \Psi_0(k)}\,,
\end{equation}
where ${\bf\Psi_0(k)}$ and ${\bf\Psi_1(k)}$ are the real and imaginary
parts of ${\bf\Psi(k)}$, respectively. It is interesting to note that
the formulae for the Hamiltonian force are very similar to the
gravitational force equation in Fourier space if we consider ${\bf
  \Psi}^{\ast}({\bf k})$ as the mass density field.

\subsection{Hamiltonian Mass and other Adjustable Parameters}
\label{sec_Hmass}

The method described above has two free parameters; the
Hamiltonian masses $m_j({\bf k})$ and the pseudo time $T$. The
efficiency of the HMC method is strongly dependent on the choices
for these parameters. Taylor et al. (2008) suggested that the
Hamiltonian mass for a variable is taken to be inversely
proportional to the width of the target probability distribution
but the suggestion made in Hanson (2001) was almost the opposite.
In this paper we take a different and much simpler approach.

When the Hamilton equations are evolved to a pseudo time $T$, to first
order approximation the change in $\delta_j({\bf k})$ is
\begin{equation}\label{eq_nm1}
\Delta\delta_j({\bf k})\simeq \frac{T}{m_j({\bf k})}p_j({\bf k})
-\left[\frac{2\delta_j({\bf k})}{P_{\rm lin}(k)}+F_j({\bf
k})\right]\frac{T^2}{m_j({\bf k})}\,.
\end{equation}
To ensure the convergence of the Hamiltonian system,
$\Delta\delta_j({\bf k})$ cannot be much larger than
$\delta_j({\bf k})$. We thus require that both the first and
second terms of $\Delta\delta_j({\bf k})$ be of the same order as
or less than $\sqrt{P_{\rm lin}(k)/2}$, the root mean square (RMS)
of $\delta_j({\bf k})$. Let us first consider the second term.
Supposing $T\sim 1$, one can deduce that the mass is of the same
order as (or less than) $\delta_j({\bf k})/(P_{\rm
lin}(k)/2)^{3/2}+F_j({\bf k})/\sqrt{P_{\rm lin}(k)/2}$. We
therefore define the Hamiltonian mass as,
\begin{equation}\label{eq_Hmass}
m_j({\bf k}) \equiv m(k) = \frac{2}{P_{\rm lin}(k)} +
\sqrt{\frac{\sum_{j=0}^1\langle F^2_j({\bf k})\rangle_{\bf
k}}{P_{\rm lin}(k)}}\,,
\end{equation}
where $\langle\cdot\cdot\cdot\rangle _{\bf k}$ denotes average
over the phase of ${\bf k}$. Note that the first and second terms in
the mass equation are actually the RMS of $\delta_j({\bf
k})/(P_{\rm lin}(k)/2)^{3/2}$ and $F_j({\bf k})/\sqrt{P_{\rm
lin}(k)/2}$, respectively. Since the momentum $p_j({\bf k})$
follows a Gaussian distribution with variance $m_j({\bf k})$, i.e.
$p_j({\bf k}) \sim \sqrt{m_j({\bf k})}$, our mass definition also
ensures that the first term in Eq (\ref{eq_nm1}) is comparable to
or less than $\sqrt{P_{\rm lin}(k)/2}$. For consistency, we set
$n_{\rm max}=13$ and choose $\tau_{\rm max}$ around $0.1$ to
guarantee that $T$ is of order unity.

The quantities $\langle F^2_j({\bf k})\rangle _{\bf k}$ vary
significantly before the HMC chain converges, and so it is not
necessary to compute the masses at every step. In practice we only
calculate the masses twice during the whole sampling. The first
calculation is before the generation of the first sample. After
proceeding $N_{\rm m}$ accepted chain steps, we use the new
Hamiltonian forces to update the mass variables and then retain
the masses all the way to the end of the sampling. In the next
section, we will show that the parameters $\tau_{\rm max}$ and
$N_{\rm m}$ have no important impact on our final results.

\subsection{Summary of Method}

Given the complicated, technical nature of the method described
above, this subsection gives a step-by-step description of the
Hamiltonian Monte Carlo method used to reconstruct the linear
density field, given a present-day density field, $\rho_{\rm
p}({\bf x})$. This serves as a `road-map' for anyone who wishes to
implement this powerful method.
\begin{enumerate}
\item Pick a cosmology, which sets the analytical, linear power
         spectrum, $P_{\rm lin}(k)$.
\item Randomly pick an initial guess for the modes $\delta({\bf k})$
         of the initial density field by specifying the corresponding real
         and imaginary parts.
\item Pick a set of Hamiltonian masses, $m_j({\bf k})$ using
         Eq.~(\ref{eq_Hmass}).
\item Randomly draw a set of momenta, $p_j({\bf k})$, from
         multi-dimensional, un-correlated Gaussian distributions with
         variances $m_j({\bf k})$.
\item Randomly pick values for the number of time steps, $n$, and the
         leapfrog time steps $\tau$, from uniform distributions in the range
         $[1,n_{\rm max}]$ and $[0,\tau_{\rm max}]$, respectively.
         In this paper, we set $n_{\rm max}=13$ and $\tau_{\rm max}=0.1$
         unless otherwise specified.
\item Integrate the Hamiltonian ``equations of motion" using the
         leapfrog technique (Eqs.~[\ref{eq_lf1}]-[\ref{eq_lf3}]) for a total
         pseudo-time $T = n\,\tau$, starting from
        $[\delta_j({\bf k}), p_j({\bf k})]$. The detailed numerical operations
         performed for each time step $\tau$ are listed below.
\item Accept the new `state' $[\delta'_j({\bf k}), p'_j({\bf k})]$, to
         which the system has evolved, with a probability given by Eq.~(\ref{paccept}).
\item Go back to step 4 and repeat until the Markov Chain has
         converged and accumulated the required number of chain elements.
\end{enumerate}

Finally, we list the numerical operations performed in each leapfrog
time step $\tau$:
\begin{itemize}
   \item  Start from the modes, $\delta_j({\bf k})$, use
    Eq.~(\ref{eq_disp}) to compute the displacement field, ${\bf s(q)}$,
    and move particles initially located on a uniform rectangular grid
    of positions ${\bf q}$ to ${\bf x(q)}={\bf q}+{\bf s}({\bf q})$.
\item Construct the MZA density field utilizing the clouds-in-cells
    (CIC) assignment method of Hockney \& Eastwood (1981).
\item Fourier transform this density field using the Fast Fourier
      Transform method, and multiply the result with the density transfer
      function $R_{\delta}(k) = \exp[0.58 \delta^2(k/2)]$ and the Gaussian
      kernel $w_{\rm G}(R_{\rm s} k)$. Divide the result with the
     Fourier transform of CIC kernel and obtain
     the modeled density field, $\rho_{\rm mod}({\bf k})$.
\item Use $\rho_{\rm mod}({\bf k})$ to compute
    $\rho_{\rm d}({\bf k})$ as described in Eq. (\ref{eq_pi1}), and
     use the method described in the paragraph below
     Eq. (\ref{eq_dvf}) to compute the density-vector field ${\bf \Psi}({\bf q})$.
\item Fourier transform ${\bf \Psi}({\bf q})$ to get ${\bf
      \Psi}({\bf k})$, and compute the likelihood term of the
      Hamiltonian forces using Eqs. (\ref{eq_f0}) and (\ref{eq_f1})
      for the real and imaginary parts, respectively.
\item Use the Hamiltonian forces to evolve the system
     according to Eqs.~(\ref{eq_lf1})-(\ref{eq_lf3}).
\end{itemize}

The main purpose of using the Gaussian kernel is to suppress
noises on the grid size $L/N_{\rm c}$ that affect the efficiency
of the HMC. We find that using $R_{\rm s}\leq 2L/N_{\rm c}$
results in a quite low acceptance rate, and so we adopt $R_{\rm
s}\geq 3L/N_{\rm c}$ throughout the paper. The efficiency of our
HMC method depends also on the value of $\mu$, in the sense that a
smaller $\mu$ leads to a lower acceptance rate. In fact, the value
of $\mu$ affects the HMC in a similar way to the smoothing scale
$R_{\rm s}$, as we will see in the next section. Moreover, in
order to achieve a good performance, the mass variables and the
pseudo-time $T=n\,\tau$ should not be specified independently but
according to their combinations in Eq. (\ref{eq_nm1}). In this
paper, we always choose $T$ to be of order unity and derive the
mass variables accordingly. Finally we note that our method is
very fast. The computations shown below are all performed using
one single processor (AMD Opteron 8380, 2.5 GHz). Each chain step
takes about 21, 222 and 2,080 seconds for $N_{\rm c}=128$, $256$
and $512$, respectively (see Table~\ref{tab_hmc}).

\section{Test with $N$-body Simulations}\label{sec:sim}

In this section we use the ``Millennium Simulation'' (MS, Springel
et al. 2005) to test our method and tune the adjustable parameters
when necessary. This simulation adopts a spatially-flat
$\Lambda$CDM model, with $\Omega_{\rm m} = 0.25$, $\Omega_{\rm b}
= 0.045$, $h = 0.73$ and $\sigma_8 = 0.9$, where $h$ is Hubble
constant and $\sigma_8$ is the {\it RMS} amplitude of linear mass
fluctuations in a sphere of of $8\mpc$ radius. It follows the
evolution of the density field with $2160^3$ particles, each
having a mass of $8.6 \times 10^8 \msun$, in a cubic box of $L =
500 \mpc$. We divide the simulation box into $N_{\rm c}^3$ grid
cells and use a Gaussian kernel with a smoothing scale of $R_{\rm
s}$ to smooth the particle distribution on to the grid (see
Table~\ref{tab_hmc} for the values of $N_{\rm c}$ and $R_{\rm s}$
used). The method used to sample the density field on the grid is
the same as that used to calculate $\rho_{\rm mod}({\bf x})$
except that now we do not include the density transfer function.
The resultant density field, denoted by $\rho_{\rm f}({\bf x})$,
is what we want to match in the reconstruction ($\rho_{\rm
p}\equiv\rho_{\rm f}$), and we assume $\sigma_{\rm p}({\bf x}) =
\mu \rho_{\rm f}({\bf x})$ as discussed in \S\ref{sec_goal}.
Moreover $\omega({\bf x})$ is always set to be unity and the
non-linear scale $k_{NL}$ used in the transfer function for the
Zel'dovich approximation is chosen to be $0.28 \mpci$.

\begin{figure*}
\epsfig{file=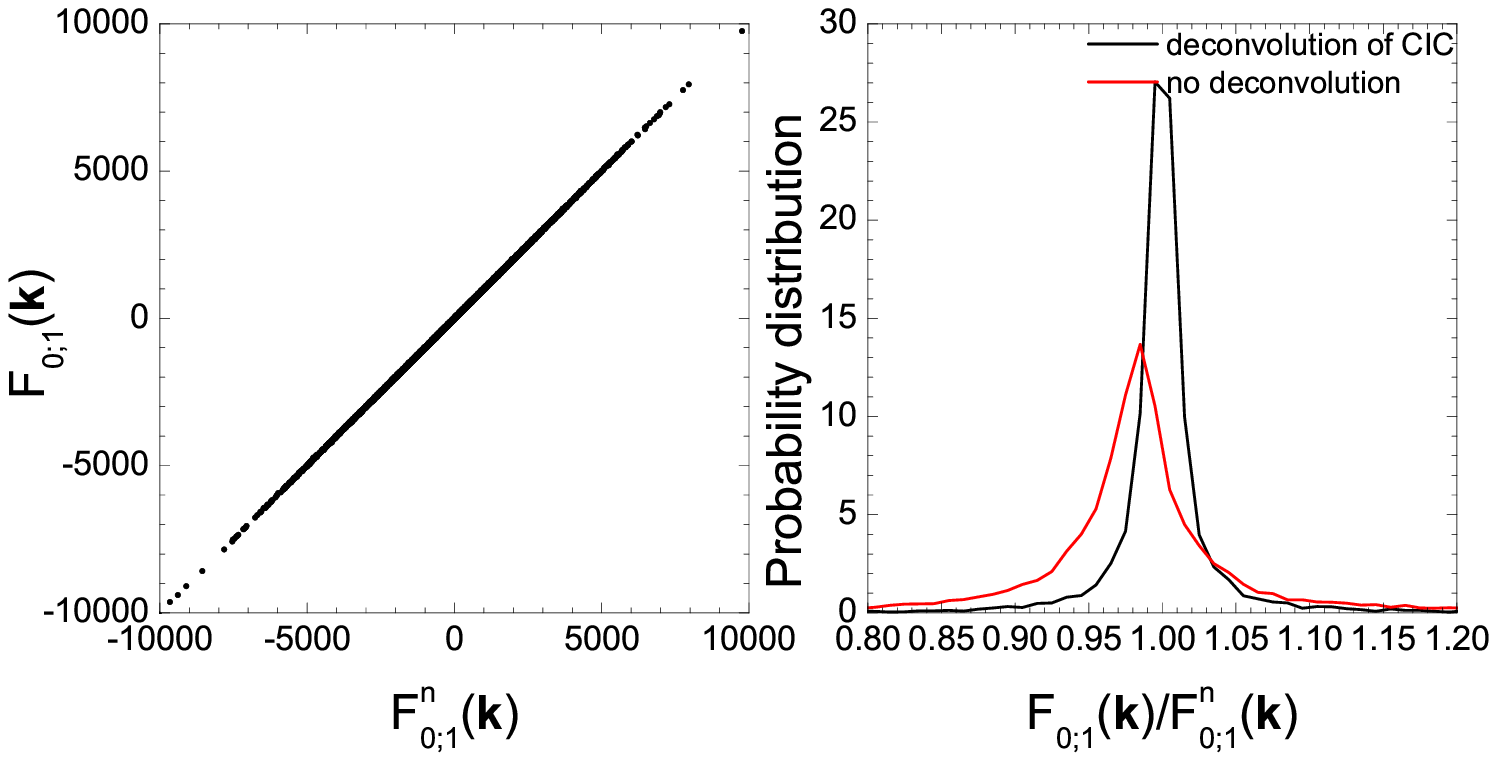,scale=1.} \caption{Comparison between the
Hamiltonian forces $F_{0;1}({\rm k})$ and $F^n_{0;1}({\rm k})$.
$F_{0;1}({\rm k})$ is calculated using Eq. (\ref{eq_f0}) and
(\ref{eq_f1}), while $F^n_{0;1}({\rm k})$ is obtained using a
numerical method presented in Section \ref{sec:sim}. The subscript
0 and 1 denote the real and imaginary parts respectively.}
\label{fig_hfc}
\end{figure*}

Before showing the test results, we verify the accuracy of our
estimation of the Hamiltonian forces. To this end, let us start with
how the forces should be calculated based on their definitions.
Suppose we want to calculate the Hamiltonian force for a chosen
variable $\delta_j({\bf k})$. We alter $\delta_j({\bf k})$ by a small
amount, $\Delta{\delta_j({\bf k})}$, with all other variables held
fixed. This leads to a small variation, $\Delta\chi^2$, in the
parameter $\chi^2$. Consequently we can obtain the corresponding force
numerically, $F^n_{j}({\bf k})=\Delta\chi^2/\Delta{\delta_j({\bf
    k})}$. This is what we would like to have. Unfortunately, this
method is very time-consuming and cannot be used to evolve the
Hamiltonian system. The left panel of Fig. \ref{fig_hfc} shows a plot
of $F_{j}({\bf k})$, obtained using Eqs.(\ref{eq_f0}) and
(\ref{eq_f1}), versus $F^n_{j}({\bf k})$. There is almost no visible
difference between these two quantities. We also present a probability
distribution of $F_{j}({\bf k})/F^n_{j}({\bf k})$ in the right panel
of Fig.  \ref{fig_hfc}. The distribution exhibits a high peak at unity
with a shallow but broad wing. Our further check finds that the
Hamiltonian forces in the broad wing are very small, which explains
why we cannot see any scatter in the left panel. We also show the
Hamiltonian forces without deconvolving the CIC kernel in deriving the
density-vector field ${\bf \Psi}({\bf q})$. The resultant $F_{j}({\bf
  k})$ is systematically smaller than $F^n_{j}({\bf k})$ due to the
smoothing with the CIC kernel. This demonstrates that our
estimates of the Hamiltonian forces based on Eqs.(\ref{eq_f0}) and
(\ref{eq_f1}) are accurate, and that deconvolution of the CIC
kernel is essential.

\begin{figure*}
\epsfig{file=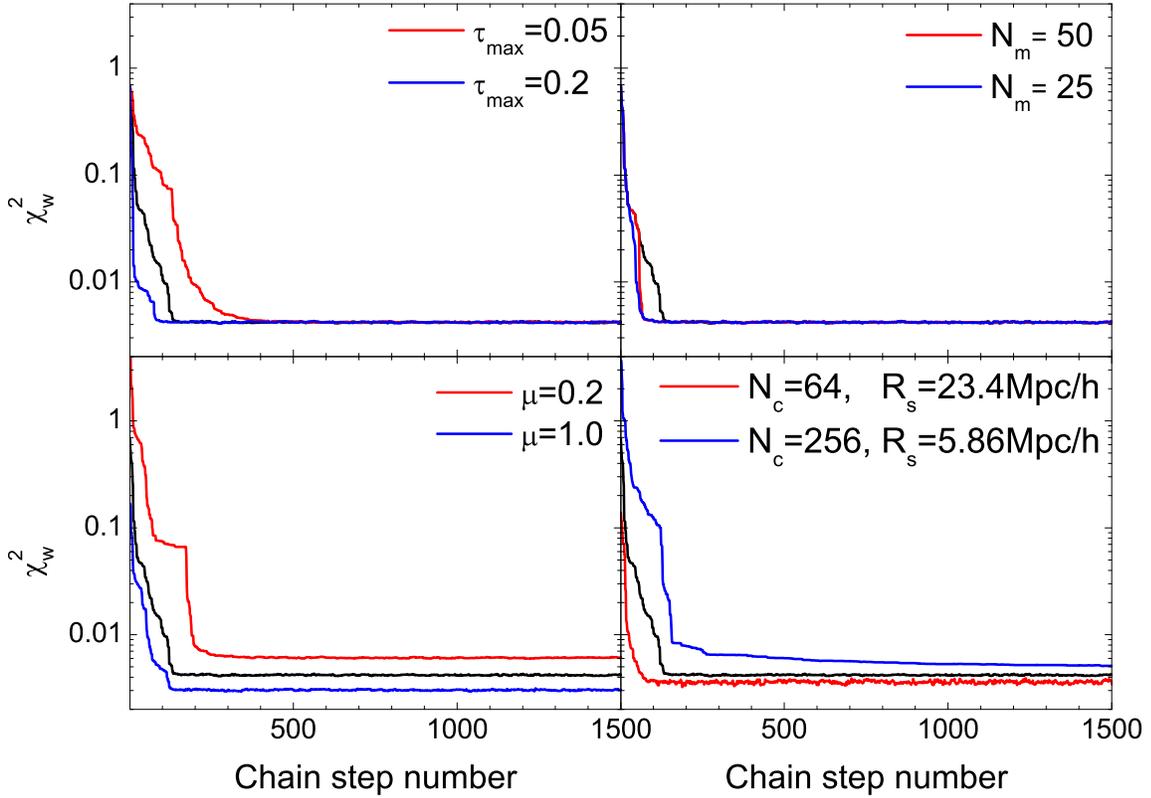,scale=1.} \caption{Evolution of $\chi^2_w$
for various tests. See text (Section \ref{sec:sim}) for the
definition of $\chi^2_w$. The black lines in the four panels are
the same, the result of primary test with parameters, $N_{\rm
c}=128$, $R_{\rm s}=11.7\mpc$, $\mu=0.5$ and $\tau_{\rm max}=0.1$.
The labels in each panel show the parameters \emph{different} from
the primary test for the corresponding test. In order to
show the burn-in phase clearly, we omitted the last 500
steps.} \label{fig_ins}
\end{figure*}

Now we apply our Hamiltonian method to the MS simulation. We first
perform a test with the following parameters: $N_{\rm c}=128$,
$R_{\rm s}=11.7\mpc$, $\mu=0.5$ and $\tau_{\rm max}=0.1$. In what
follows this test is referred to as ``primary test". The initial
set of $\delta_j({\bf k})$ is randomly drawn from the prior
probability distribution given by Eq.(\ref{eq_gs}). We make
calculations of 2,000 HMC steps and the average acceptance rate is
$A_r=72\%$. Fig.\ref{fig_ins} shows $\chi^2_w=\chi^2/\sum_{\bf
x}\omega({\bf x})$ (here, $\sum_{\bf x}\omega({\bf x})=N_{\rm
c}^3$) as a function of chain step. One can see that $\chi^2_w$
drops sharply at the beginning (the burn-in phase), then remains
almost constant after about 150 chain steps (the convergence
phase). The density field of a converged chain step matches well
the original input density field, with a RMS difference between
the two about $\mu\sqrt{2\chi^2_w}\simeq 4.6\%$. For reference,
the important parameters and characteristics of the primary test
are listed in Table \ref{tab_hmc}.

In Fig. \ref{fig_evps}, we show the power spectra measured from
different chain steps. We refer these power spectra as the Hamiltonian
power spectra [$P_{\rm H}(k)$], to distinguish them from the input
linear power spectrum and the analytic linear power spectrum. The
`two-phase' behavior can be clearly seen in the power spectra
evolution. For the first 150 steps, one sees an obvious tuning
process: on large scales the Hamiltonian spectrum first decreases and
then increases, while on small scales the behavior is the
opposite. The first (starting) spectrum is very similar to the
analytic linear power spectrum, because the initial $\delta_j({\bf
  k})$ are drawn from Eq.(\ref{eq_gs}). After about 150 steps, the
Hamiltonian spectra settle close to the initial spectrum used in the
simulation, implying the convergence of the chain.  Both the $\chi^2$
and the power spectrum results demonstrate that our method ensures
quick convergence.

\begin{table}
\begin{center}
\caption{The important parameters and characteristics for HMC and
resimulation. Here $L$ and $R_{\rm s}$ are in unit of $\mpc$, while the
softening length $\epsilon$ is in unit of $\kpc$. The mean
consumption time $t_c$ for each chain step is in second. $A_r$ is
the acceptance rate. The particle mass $m_p$ for resimulation is
in unit of $10^{10}\msun$. }\label{tab_hmc}
\begin{tabular}{lccccc}
  \hline\hline
  ~~~~~~~~~~~~~~~~~~~~~~~~~~~~~~~~~~~~~~~~~& HMC & ~~~~~~~~~~~~~~~~~~~~~~~~~~~~~~~~~~ & Resimulation &~~~~~\\
\end{tabular}
\begin{tabular}{lccccccccccccccccccccccccccccccc}
  \cline{2-7}\cline{9-12}
     & $L$ & $N_{\rm c}$ & $R_{\rm s}$ & $\chi^2_w(10^{-3})$ & $t_c$ & $A_r$ & & $N_p$ & $m_p$ & $\epsilon$ & $z_i$ \\
  \hline
  Primary Test & 500 & 128 & 11.7 & 4.2 & 21  & $72\%$ & & $128^3$  & 414  & 80  & 20\\
  LRR & 726 & 512 & 5.67 & 2.9 & 2080 & $35\%$ & & $512^3$  & 19.8  & 28 & 30\\
  HRR & 181.5 & 256 & 2.84 & 3.8 & 222  & $58\%$ & & $256^3$  & 2.47  & 15 & 36\\
  \hline
\end{tabular}
\end{center}
\end{table}

Such two-phase behavior is common in HMC (e.g. Hanson 2001; Taylor
et al. 2008) and can be understood in terms of the behavior of a
physical system. The system initially has very large
potential and large $\chi^2$ because its initial configuration,
represented by the randomly generated initial $\delta_j({\bf k})$,
is unstable and tends to fall rapidly into the deep potential
well, which reduces $\chi^2$ and potential of the system but
increases the kinetic energy. Since the kinetic energy is
directly related to the current momenta which are reset to
lower values before each chain step, the system is `cooled'
so that it continuously falls towards the potential well
(burn in phase). Once the changes of the potential and
kinetic energy become insignificant within one chain step,
the `cooling process' can be neglected and the total energy of
the system becomes stable and so the value of $\chi^2$
remains more or less constant. In this convergence phase,
the accepted states reach the bottom region of the
potential well around the posterior peak we are searching.

Recall again that we want to generate a linear density field
which obeys the prior Gaussian probability function specified by a
linear power spectrum and evolves to a non-linear density field
that matches the input density field. In what follows, we will
examine our results in two different aspects. One is whether or
not our reconstructed $\delta_j({\bf k})$ matches the prior
constraints. The other is how well the predicted (or modeled)
density field (from the MZA), $\rho_{\rm mod}$, and the original
simulated density field, $\rho_{\rm f}$, match each other.
In the conventional application of the HMC method, one
generally performs the correlation length test
(see e.g. Taylor et al. 2008; JW13) to determine the amount of
independent samples that are drawn from a HMC chain.
However, for our purpose we are not trying to sample
the whole posterior distribution, it is not necessary
to perform the correlation analysis.

The very small $\chi^2_w$ for the converged chain states demonstrates
that our method can recover the input density field at high
accuracy. To further quantify at which scales $\rho_{\rm mod}$ matches
$\rho_{\rm f}$ well, we measure the phase correlation between the two
density fields. We define a phase correlation coefficient between two
fields $X({\bf k})$ and $Y({\bf k})$ as
\begin{equation}
C_{p}(k) = X\otimes Y = \frac{\langle X_0({\bf k})Y_0({\bf
k})+X_1({\bf k})Y_1({\bf k})\rangle _{\bf k}}{\sqrt{\langle
|X({\bf k})|^2\rangle_{\bf k}\langle |Y({\bf k})|^2\rangle_{\bf
k}}}\,,
\end{equation}
where the subscripts 0 and 1 indicate the real and imaginary
parts, respectively. Note that $C_{p}(k)=1$ means that the two
quantities have exactly the same phase, while $C_{p}(k)=0$
indicates no correlation. We show the phase correlation between
$\rho_{\rm f}$ and $\rho_{\rm mod}$ at the 1,700th chain
step for our primary test in Fig. \ref{fig_phcr}. The
correlation coefficient is close to one at large scales
($k<0.17\mpci$), and declines quickly towards smaller scales. Such
a sharp transition is due to the fact that small-scale structures
are severely suppressed by our Gaussian smoothing and so
contribute little to $\chi^2$ and to the Hamiltonian force. As a
result, the transition scale should be related to the smoothing
scale. To demonstrate this, we perform two additional tests with
$R_{\rm s}=23.4\mpc$ and $R_{\rm
  s}=5.86\mpc$, assuming the same $N_{\rm m}$, $\tau_{\rm max}$ and $\mu$ as
in the primary test. We adopt $N_{\rm c}=64$ for the test with
large $R_{\rm s}$, and $N_{\rm c}=256$ for the one with smaller
$R_{\rm s}$ one. For $R_{\rm s}=5.86\mpc$ the correlation
coefficient remains about unity until $k>0.3\mpci$, while for
$R_{\rm s}=23.4\mpc$ the transition starts already from
$k\sim0.08\mpci$. As expected, the transition scale is inversely
proportional to the value of $R_{\rm s}$. We also show $\chi^2_w$
as a function of chain steps for these two tests in Fig.
\ref{fig_ins}. Similar to the primary test, their $\chi^2_w$ first
declines rapidly and then stabilizes at some small values.

Now let us check whether our method can recover the power spectrum
and the Gaussian distribution used as priors in our
reconstruction. Inspecting the Hamiltonian spectra in more detail,
we see that, over the entire range of wavenumbers, the Hamiltonian
spectra converge to the linear power spectrum used in the MS
simulation. However, there is noticeable discrepancy between the
Hamiltonian spectra and the linear power spectrum on intermediate
scales ($\sim 0.2 h\,{\rm Mpc}^{-1}$). The distributions of
$\delta_j({\bf k})/\sqrt{P_{\rm lin}(k)/2}$ are shown in Fig.
\ref{fig_gaudis} for three different (large, intermediate and
small) scales.  These distributions can all be well fitted with a
Gaussian, as expected, with the values of $\sigma$ shown in the
corresponding panel. The best-fitting $\sigma$ on both small and
large scales are very close to the expectation value, $\sigma=1$,
indicating the spectrum is well reproduced on these scales. On
intermediate scales, however, a deviation of $\sigma$ from unity
is clearly seen, consistent with the power spectrum result.

To understand the origin of this discrepancy on intermediate
scales, we perform a series of tests with smoothing scales ranging
from $3\mpc$ to $23\mpc$. In each case, we find that the input
linear power spectrum is well recovered at both large and small
scales, but that noticeable discrepancies are apparent on
intermediate scales. The discrepancy moves gradually from small to
large wavenumbers as the smoothing scale decreases (see e.g. the
lower right panel of Fig. \ref{fig_dpps}). We define a wavenumber
$k_{\rm c}=\sum_{k}k(1-r(k))/\sum_k(1-r(k))$ with $r(k)=P_{\rm
H}(k)/P_{\rm lin}(k)$ to quantify this scale, and find that the
$k_{\rm c}$ - $R_{\rm s}$ relation can be well fitted by $k_{\rm
  c}=1.88/R_{\rm s}^{0.94}$. Note that the phase correlation between
$\rho_{\rm mod}$ and $\rho_{\rm f}$ also depends on the smoothing
scale. We show $k_{\rm c}$ so defined as the dashed lines in the phase
correlation plot (Fig. \ref{fig_phcr}), which clearly shows that
$k_{\rm c}$ also characterizes the transition scale in the phase
correlation. The question is why the reconstructed linear power
spectrum has the deviation from the analytical linear power spectrum
just around $k_{\rm c}$. According to Eq.(\ref{eq_dp}), the
Hamiltonian force has two components, the prior term and the
likelihood term. The phase can be well recovered only when the
likelihood term dominates the force because the prior term actually
generates random phases. The mean ratio of these two terms is,
\begin{equation}
R_F(k)= \frac{P_{\rm lin}(k)}{2}\sqrt{\frac{\sum_{j} \langle F^2_j({\bf
k})\rangle_{\bf k}}{\sum_{j}\langle\delta^2_j({\bf k})\rangle_{\bf k}}}\,.
\end{equation}
We find that $R_F(k)$ decreases continuously with $k$ and is about
unity around $k_{\rm c}$. This implies that the $\chi^2$ is more
sensitive to $\delta({\bf k})$ at smaller $k$, and is almost
completely independent of $\delta({\bf k})$ at $k\gg k_{\rm c}$. At
$k\gg k_{\rm c}$, the trajectories of $\delta({\bf k})$ in the HMC are
dominated by the prior so that the Hamiltonian spectra match the
linear power spectrum but the Fourier phases are not constrained. At
$k\ll k_{\rm c}$, on the other hand, the trajectories of $\delta({\bf
  k})$ is governed by the likelihood term so that they try to trace
the original linear density field of the MS simulation, consequently
leading to a small $\chi^2$ and a strong phase correlation between the
reconstructed and original density fields. However, on scales around
$k_{\rm c}$, $\delta({\bf k})$ has a reduced contribution to $\chi^2$
and the posterior distribution is partly regulated by the prior.
Consequently the final result is a compromise between the prior
constraint and the likelihood. Since the likelihood term decreases
with increasing smoothing scale [see Eqs.(\ref{eq_dvf}), (\ref{eq_f0})
and (\ref{eq_f1})] while the prior term is not, it explains why both
$k_{\rm c}$ and the transition scale of the phase correlation depend
linearly on the smoothing scale.

Although the discrepancy between the reconstructed and original
linear power spectra is not big, we may want to correct it using a
`renormalization' process. First we visually identify the region
where the discrepancy is significant ($0.12<k<0.21\mpci$ for the
primary test), and then use $\sqrt{P_{\rm lin}(k)/P_{\rm H}(k)}$
to scale the amplitude of $\delta_j({\bf k})$ without changing its
phase. Since $\delta_j({\bf k})$ in this region are still well
described by a Gaussian distribution (see Fig. \ref{fig_gaudis}),
and since rescaling preserves the shape of the distribution, the
distribution of the scaled $\delta_j({\bf k})$ is still close to
Gaussian. Because the discrepancy to be corrected is fairly small
and the contribution of $\delta({\bf k})$ around $k_{\rm c}$ to
$\chi^2$ is not large, the scaling does not cause any significant
change in $\chi^2$ and the phase correlation. In Section
\ref{sec:resim}, we will use the renormalized $\delta({\bf k})$ to
generate initial conditions at high redshift, then use a $N$-body
simulation to evolve the initial condition to the present day and
compare the re-simulated density fields with the original simulated
density field.

Finally, we discuss the effects of changing $\tau_{\rm max}$,
$N_{\rm m}$ and $\mu$ on our results. We perform several tests
with different $\tau_{\rm max}$, $N_{\rm m}$ and $\mu$. Similar to
the primary test, all these tests exhibit a two-phase behavior,
and a final convergence with a low $\chi^2_w$ is always achieved
(see Fig. \ref{fig_ins}). Inspecting the results in detail, one
can see that the values of $\tau_{\rm max}$ and $N_{\rm m}$ affect
the number of steps required for burn-in: large $\tau_{\rm max}$
and small $N_{\rm m}$ both result in a quick burn-in phase.
Similar to $R_{\rm s}$, the value of $\mu$ also affects the
difference between the converged $\rho_{\rm mod}$ and $\rho_{\rm
f}$, in the sense that a smaller $\mu$ results in smaller
$\mu\sqrt{2\chi^2_w}$, as shown in the lower left panel of
Fig.\,\ref{fig_ins}. Furthermore, the value of $k_{\rm c}$
decreases with the increase of $\mu$ (see the lower left panel of
Fig. \ref{fig_dpps}), because a larger $\mu$ suppresses more the
contribution of small-scale structures to $\chi^2$. Thus $\mu$
affects the accuracy of the final result in a way quite similar to
$R_{\rm s}$. On the other hand, since the MZA becomes increasingly
inaccurate on small scales, a smaller $\mu$ also leads to a lower
acceptance rate. As a compromise between efficiency and accuracy,
and because the effects of changes in $\mu$ and $R_{\rm s}$ are
degenerate, we fix $\mu=0.5$ and only test the impact of changing
$R_{\rm s}$. We have also checked the Hamiltonian power spectra
and the probability distribution of $\delta_j({\bf
k})/\sqrt{P_{\rm lin}(k)/2}$ in these tests and found that they
are very similar to those in the primary test (Fig.
\ref{fig_dpps}).

\section{Application to Reconstructed Density Field}
\label{sec:mocktest}

As we discussed above, the HMC method needs a present-day density
field, $\rho_{\rm p}$, as an input. Therefore the method is useful
only when we have a reliable method to obtain the present-day
density field from observation. In Wang et al. (2009a), we
developed a method to reconstruct the present-day density field
based on the distribution of galaxy groups (haloes). In this
section, we apply this method to a mock group catalogue to
reconstruct the density field and then apply our HMC method to the
reconstructed density field to obtain the initial linear density
field ($\rho_{\rm p}\equiv\rho_{\rm rc}$). In the first
subsection, we briefly describe how we construct the mock galaxy
and group catalogs and the method to correct for redshift space
distortions of the groups. The two mock catalogs are exactly the
same as those used in Wang et al. (2012a) and are constructed from
the MS simulation with the use of the SDSS DR7 sky coverage and
selection functions. We refer the interested readers to Wang et
al. (2009a, 2012a) and references therein for further details.

\subsection{The Mock Galaxy and Group Catalogues}

The construction of the mock galaxy catalogue is similar to that
in Yang et al. (2007, hereafter Y07). First, we populate
galaxies inside dark matter halos according to the conditional
luminosity function (CLF, Yang, Mo \& van den Bosch 2003) model of
van den Bosch et al. (2007). These halos are identified from the
MS simulation with a friends-of-friends algorithm (hereafter FOF,
Davis et al. 1985) employing a linking length of $b=0.2$. Next, we
assign phase-space parameters to these galaxies following Yang et
al. (2004; see also More et al. 2009). Briefly, in each halo, the
brightest galaxy is regarded as the central galaxy and assumed to be
located at the halo center, while the other galaxies are
satellites and distributed spherically following an NFW (Navarro,
Frenk \& White 1997) density profile, with the concentration-mass
relation given by Macci\`o \etal (2007). The peculiar velocity of
a central galaxy is set equal to the mean velocity of the
corresponding halo, while satellites have additional velocity
components to account for the virial motions within the host halo.
This component of velocity is drawn from a Gaussian probability function with
a dispersion computed from the Jeans equation (see More et al.
2009). Then we stack $3\times3\times3$ galaxy-populated simulation
boxes together and place a virtual observer at the center of the
stack. We assign each galaxy a redshift and ($\alpha_{\rm J}$,
$\delta_{\rm J}$) coordinates with respect to the observer. Note
that the redshift of a galaxy is a combination of its cosmological
redshift and peculiar velocity. The mock galaxy catalogue is
constructed by mimicking the sky coverage of the SDSS DR7 and
taking into account the angular variation of the magnitude limit
and the survey completeness (see Li et al. 2007).

 Mock galaxy groups are identified with a halo-based group
finder developed by Yang et al. (2005). This group finder has
already been successfully applied to the SDSS DR4 (Y07). The
application to our mock galaxy catalogue is the same, except that
different cosmology and larger sky coverage are adopted. Groups
are selected in the survey region that has the redshift
completeness criterion of ${\cal C}_z> 0.7$. The masses of
groups are estimated based on the ranking of the characteristic
luminosities of groups. The characteristic luminosity of a group
is determined from the luminous group members with $M_r - 5 \log h
\geq -19.5$. To take account of the survey edge effect, the group
finder calculates the fraction of each group's volume that falls
inside the survey region, then uses this fraction to correct for
the group luminosity and mass. In this paper, we use all groups
with assigned mass $M_h\ge M_{\rm th}=10^{12}h^{-1}{\rm M}_\odot$
and we restrict our reconstruction of the present-day density
field to the cosmic volume covering the redshift range $0.01 \leq
z \leq 0.12$, which we call the survey volume.

In order to reconstruct the density field in real space, we have
to correct for redshift space distortions. To do that, we follow
exactly the same procedure as described in Wang et al. (2012a, see
also Wang et al. 2009a). First, we embed the survey volume in a
periodic cubic box of side $726\mpc$, which is referred to as
survey box in the following. We divide the survey box into
$1024^3$ grids and assign the masses of groups (with $M_h\ge
M_{\rm th}$) on the grids according to their redshift-space
coordinates. The grids outside of the survey volume are assigned
the mean mass density of the groups ($M_h\ge M_{\rm th}$) in the
survey volume. Then we calculate the overdensity field of the
groups, and smooth it using a Gaussian kernel with a large
smoothing scale of $8\mpc$. In the linear regime, the peculiar
velocities induced by density perturbations are proportional to
the amplitudes of the density fluctuations, hence we can use this
smoothed overdensity field to derive the peculiar velocities.
Adopting the relatively large smoothing scale can effectively
suppress non-linear velocities that cannot be predicted
accurately. As shown in Wang et al. (2009a), the resultant
velocities based on the halo (group) population are biased but
tightly proportional to the real velocities. We thus can predict
the peculiar velocity of each group by simply taking into account
the bias factor of the overdensity field represented by the
groups, and we use equation (10) in Wang et al. (2009a) to
calculate the bias factor.  Finally we use the line-of-sight
component of the predicted velocity to correct for the redshift
space distortions. Since the velocity field is obtained from the
groups distributed in redshift space, this method requires
iteration to make sure that convergence is achieved (typically
twice; Wang et al. 2012a).

According to the results obtained in Wang et al. (2009a) and
from our further tests, the average offset between the real
and predicted positions of groups is between $1.1$ to
$1.4\mpc$. Using the 2LPT to correct for the redshift distortion
can achieve a higher accuracy than the simple linear model
(Kitaura et al. 2012b).
However, in our case it is not necessary to adopt the more
accurate 2LPT, because the density field reconstructed from the
mock groups is to be smoothed on a scale of at least $2.84\mpc$
(see below). This smoothing scale is larger than the
typical offset so that the linear theory and the 2LPT
lead to very similar results. Another important issue related to
redshift distortion is the finger-of-god effect. This
effect is mainly due to the virial motion within
individual groups, especially massive groups, and
thus can not be handled by a method based on
linear or quasi-linear model. Such effect, if not properly
corrected, can lead to artifacts in the reconstructed
density field. An advantage in our reconstruction is
that it is based on galaxy groups rather than galaxies. The
redshift of a group is estimated using the luminosity-weighted
average of all member galaxies so that the finger-of-god effect
within the group is largely mitigated (see figure 8
in Y07 for a comparison between the distributions of
groups and galaxies). However, some small groups that
are close to massive groups may have significant
non-linear motion (see Wang et al. 2009b), and such
motion cannot be corrected properly with any
linear or quasi-linear model. Fortunately the total
mass contained in such groups is very small and
the their effect is expected to be small.

\subsection{Density Reconstruction from Mock Group Catalogue}
\label{sec_denrc}

In Wang et al. (2009a), we developed a method to reconstruct the
present-day density field starting from the distribution of galaxy
groups (haloes). We tested this method using FOF haloes but did
not apply it to mock groups to examine its reliability against
uncertainties due to false identification of group members and
survey boundary effect. Survey boundaries have no direct impact on
our reconstruction, but can affect the correction for redshift
space distortions.  In this subsection, we reconstruct the density
field using the mock group catalogue corrected for redshift space
distortions as described above and compare our reconstruction with
the original simulation.

As in Wang et al. (2009a), we first calculate the density profiles
in `domains' using the MS simulation. For a given population of
FOF haloes with mass of $M_{\rm h}\geq M_{\rm th}$, we partition
the cosmic space into a set of domains in such a way that each
domain contains one and only one halo. Any point in the domain of
a halo is closer (according to a distance proximity defined below)
to this halo than to any other haloes. The proximity of a point to
a halo with viral radius of $R_h$ is defined as,
\begin{equation}
r_p=\frac{r_h}{R_h}\,,
\end{equation}
where $r_h$ is the physical distance from the point to the halo
center. We calculate the average density profile around halos of
the same mass in the corresponding domains. Fig. \ref{fig_dp12}
shows the results in six different mass bins. Despite using a
different cosmology, these profiles are very similar to those
shown in Wang et al. (2009a). Then we `convolve' our mock groups
with these density profiles to reconstruct the cosmic density in
the following way. For a mock group of mass above the mass
threshold $M_{\rm th}$, we pick a density profile shown in Fig.
\ref{fig_dp12} according to the mass of this group. Using Monte
Carlo method, we put particles around this group up to $\sim 32$
times the virial radius regardless of the domain. We remove
particles outside of the domain of the group. We repeat the above
three steps for all groups with mass larger than $M_{\rm th}$.
Eventually we get a present-day density field, embedded in the
survey box.

One advantage of our reconstruction method is that we define a
very special cross correlation function  (i.e. the average density
profile in domain). It is different from the conventional cross
correlation, which does not use domain. In the conventional one,
all haloes contribute to the average density profiles at all
scale, especially on large scales. This has the effect of
smoothing halo masses over very large scales. On the other hand,
in our cross correlation based on domains, massive haloes ($M_{\rm
h}\geq M_{\rm th}$) contribute only to the density profiles within
the virial radii, while the diffuse mass and low-mass halos with
$M_{\rm h}<M_{\rm th}$ contribute to the large scale. Another
advantage of our method is that we use galaxy groups (haloes)
instead of galaxies. While the bias of galaxy distribution
relative to the underlying density field may depend on various
galaxy properties, such as luminosity and color, and the exact
form of the bias is not well established, the use of galaxy groups
(haloes) automatically takes into account the bias of the galaxy
distribution through the connection between galaxies and haloes.

In order to compare our reconstructed density field with the
simulated density field, we divide the survey box into $1024^3$
grids with size of $0.71\mpc$, and smooth the sampled particles on
the grids using a Gaussian kernel with $R_{\rm s}=2.84\mpc$. Since
the peculiar velocities are predicted more accurately in the inner
region than near the boundary of survey volume (Wang et al. 2012a)
and we use these velocities to correct for the redshift space
distortions, one may expect that the reconstruction is more
reliable in the inner region. To check whether or not this is the
case, we need to compute the distance of any point in the survey
volume to the boundary. Following Wang et al. (2012a), we define a
filling factor $p_f$, to characterize the closeness of a grid to
the boundary. For a grid ($g$) in the survey volume, $p_f$ is the
fraction of a spherical volume of radius $80\mpc$ centered on the
grid $g$, that falls inside the survey volume. Therefore, $p_{\rm
f} \simeq 1$ for grids located more than $80\mpc$ from any survey
boundary, while it is much less than unity for grids close to the
boundary.

In Fig. \ref{fig_rhorc} we present the comparison between the
simulated density field $\rho_{\rm f}$ and the reconstructed
density field $\rho_{\rm rc}$ in the inner region with $p_{\rm
f}\geq0.6$, which is about $66\%$ of the survey volume. The solid
line shows the mean relation and the error bars indicate the
standard deviation in $\rho_{\rm f}$ for a given $\rho_{\rm rc}$.
The bias is very small and the uncertainties are about $30\%$ to
$50\%$ of $\rho_{\rm rc}$ in most bins. At the two largest density
bins, there is a significant deviation from the one to one
relation. The volume of the grids in these two bins is tiny and
their presence does not affect our reconstruction significantly.
We then show the same comparison for grids near the boundary, i.e.
with $p_{\rm f}<0.6$. As one can see, the result is as good as
that in the inner region. Apparently, the smoothing used is able
to remedy partly the problem in the correction of the redshift
space distortions near the survey boundary. Overall, with an
appropriate choice of the smoothing radius, our method is able to
reconstruct the density field accurately, and the effects due to
survey boundary and group contamination do not introduce any
significant bias.

In the next subsection, we will apply the HMC method to the
reconstructed density field in two volumes, one is the entire
survey volume and the other is a cubic volume well inside the
survey volume (see below for details). Before doing that, we show
in Fig.\ref{fig_phrs} the phase correlation between our
reconstructed and simulation fields in these two volumes.  These
correlations set an upper limit on the accuracy we can achieve for
our reconstruction of the linear density field. As one can see,
the correlation coefficient is almost unity on large scales, and
declines gradually with the increase of the wavenumber. At
$k\sim1\mpci$ the coefficient drops to $\sim 0.5$, demonstrating
that our method is successful well into the nonlinear regime (the
translinear scale is $k\sim 0.15 \mpci$). The phase correlation at
$k>0.2\mpci$ obtained for the inner cube is slightly stronger than
that for the total volume, indicating that boundary effects indeed
have a non-negligible impact on the reconstruction.

\subsection{Reconstruction of Linear Density Field}

We first apply our HMC method to the reconstructed density field
in the entire survey volume, which is embedded in the survey box
with size $L=726\mpc$ on a side. In order to examine the ability
of our method over a large dynamical range, we divide the survey
box into $N_{\rm c}=512$ grids in each dimension, so that our
reconstruction deals with more than $N_{\rm c}^3\sim 10^8$ free
parameters. We adopt a smoothing radius of $R_{\rm s}=5.67\mpc$,
on which our reconstructed density field is quite similar to the
original density field. In order to derive the weight $w({\bf
x})$, we divide each grid into $2^3$ subgrids. If more than 6
subgrids of a grid are located inside the survey volume, this grid
is assigned a weight of unity, otherwise zero. The other
parameters are chosen as $\tau_{\rm max}=0.1$, $\mu=0.5$ and
$N_{\rm m}=50$, similar to those in the primary test. This
application is referred to as the `Low Resolution Run' or LRR in
the following.  As shown in Table~\ref{tab_hmc}, on average it
takes $\sim 2080$ seconds for each HMC step and the acceptance
rate is about $35\%$. The $\chi^2_w$ value for a converged state
is about 0.0029, indicating that the RMS difference between
$\rho_{\rm
  mod}({\bf x})$ and $\rho_{\rm rc}({\bf x})$ is only 3.8\%.  We show
the phase correlation between $\rho_{\rm rc}$ and $\rho_{\rm mod}$
at the 1700th chain step in the left panel of Fig.
\ref{fig_phrs}. The correlation is almost unity at large scales,
and drops quickly around $k_{\rm c}$, consistent with our test
results presented above.

In the left panel of Fig. \ref{fig_ps} we compare a converged
Hamiltonian spectrum, $P_{\rm H}(k)$, with the analytic linear
power spectrum, $P_{\rm lin}(k)$, used in the prior.  As one can
see, $P_{\rm H}(k)$ is very close to $P_{\rm lin}(k)$ over the
entire range of scale. In particular, the discrepancy around
$k_{\rm c}$ is also small. The distribution of the reconstructed
initial density field is extremely close to Gaussian, as shown in
the upper three panels of Fig. \ref{fig_mgau}. As described above,
we can also renormalize the reconstructed linear power spectrum by
visually identifying the discrepancy region {\bf
($0.28<k<0.47\mpci$)} and scaling the corresponding $\delta_j({\bf
k})$ with a factor $\sqrt{P_{\rm lin}(k)/P_{\rm H}(k)}$. The new
power spectrum so obtained and the distributions of the
re-normalized $\delta_j({\bf k})$ are also shown in the
corresponding figures. Since the discrepancy is tiny, the curves
before and after the renormalization are undistinguishable in the
figure. As shown in the upper panels of Fig. \ref{fig_mgau}, these
re-normalized $\delta_j({\bf k})$ are well described by a Gaussian
function with unity variance. These results demonstrate clearly
that our HMC method works well on large scales.

To examine the performance of our HMC method on small scales, we
perform a High Resolution Run (HRR) with a small smoothing scale.
This is a cubic box of side $100\mpc$ located in the inner region
of the survey volume, put inside a larger periodic box with
$L=181.5\mpc$. We divide the larger box into $N_{\rm c}=256$ grids
in each dimension, and adopt a smoothing scale of $R_{\rm
s}=2.84\mpc$, for which our reconstructed density field is in good
agreement with the original simulation. Only grid cells that are
located within both the small box and the survey volume are
assigned a weight of unity. All other grid cells are assigned a
weight of zero. Note that some grids in the small box may not be
in the survey volume, because of the existence of small holes in
the survey mask. The other parameters are chosen to be the same as
in the LRR. On average the HRR takes 222 seconds for each HMC
step, and the acceptance rate is about $58\%$. The chain finally
converges to $\chi^2_w\simeq 0.0038$, corresponding to a RMS of
$4.4\%$ in the difference between $\rho_{\rm
  mod}({\bf x})$ and $\rho_{\rm rc}({\bf x})$. The corresponding phase
correlation is shown in the right panel of Fig. \ref{fig_phrs}. Here
we see a significant correlation all the way to $k\sim 1 \mpci$, a
scale much smaller than the translinear scale which corresponds to
$k\sim 0.15 \mpci$.

The converged $P_{\rm H}(k)$ and the distributions of
$\delta_j({\bf k})/\sqrt{P_{\rm lin}(k)/2}$ for the HRR are shown
in Figs. \ref{fig_ps} and \ref{fig_mgau}. Here we see a
significant bump at $0.3<k<1\mpc$ $P_{\rm H}(k)$ compared with
$P_{\rm lin}(k)$. The reason for this bump is twofold. First, the
bump is around $k_{\rm c}$, suggesting that it is a compromise
between the prior constraint and the likelihood (see Section
\ref{sec:sim}). Second, the bump is partly ascribed to the
inaccuracy in the adopted approximation of the structure formation
model on small scales. According to TZ12, the mode-coupling term,
$\rho_{MC}({\bf k})$, cannot be neglected at $k>0.4\mpci$ (see
their figure 1). On such scales, the amplitude of $\rho_{\rm
mod}({\bf k})$ predicted by the MZA is somewhat lower than that of
the fully evolved density field. To achieve a small $\chi^2$ in
the HMC, the Hamiltonian spectra have to be enhanced to compensate
the deficit. On even smaller scales, $k> 1\mpci$, however, the
Hamiltonian force is dominated by the prior term so that $P_{\rm
H}(k)$ is forced back to $P_{\rm lin}(k)$. Despite of this
inaccuracy, our HMC method can still recover more than half of the
phase correlation all the way to $k\sim 1\mpci$, as demonstrated
in the next section. As before, we re-normalized the amplitudes of
the Fourier modes to correct for the discrepancy in the range of
$0.35<k<1\mpci$. The corrected power spectrum and distribution of
the re-normalized $\delta_j({\bf k})$, are shown in the right
panel of Fig. \ref{fig_ps} and the lower panels of
Fig.\ref{fig_mgau}, respectively.

\begin{figure*}
\epsfig{file=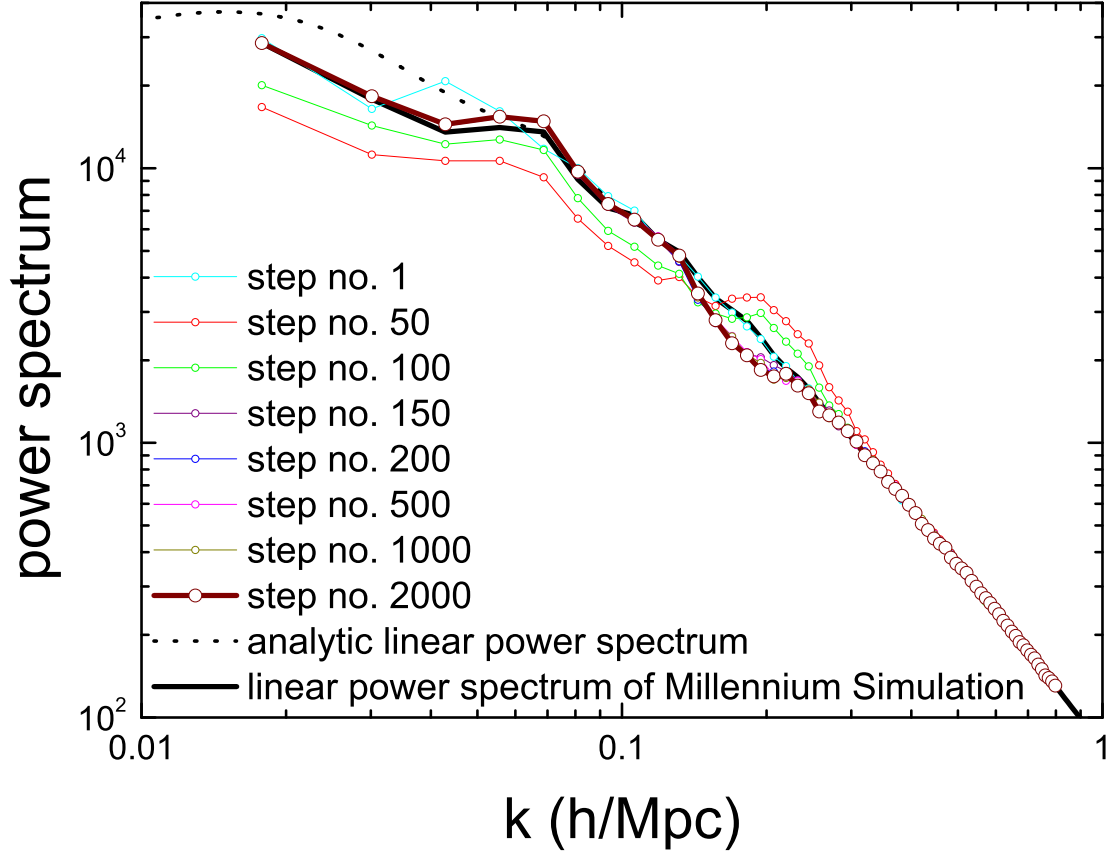,scale=1.} \caption{Circles connected by
lines show the evolution of Hamiltonian power spectra, $P_{\rm H}(k)$,
at different chain steps for the primary test. Dotted line shows
the analytic linear power spectrum, $P_{\rm lin}(k)$, and the black solid
line shows the power spectrum measured from the MS simulation. In
order to clearly compare the convergent $P_{\rm H}(k)$ with $P_{\rm lin}(k)$ and
the MS spectrum, we show $P_{\rm H}(k)$ at the 2000th step in bold.}
\label{fig_evps}
\end{figure*}

\begin{figure*}
\epsfig{file=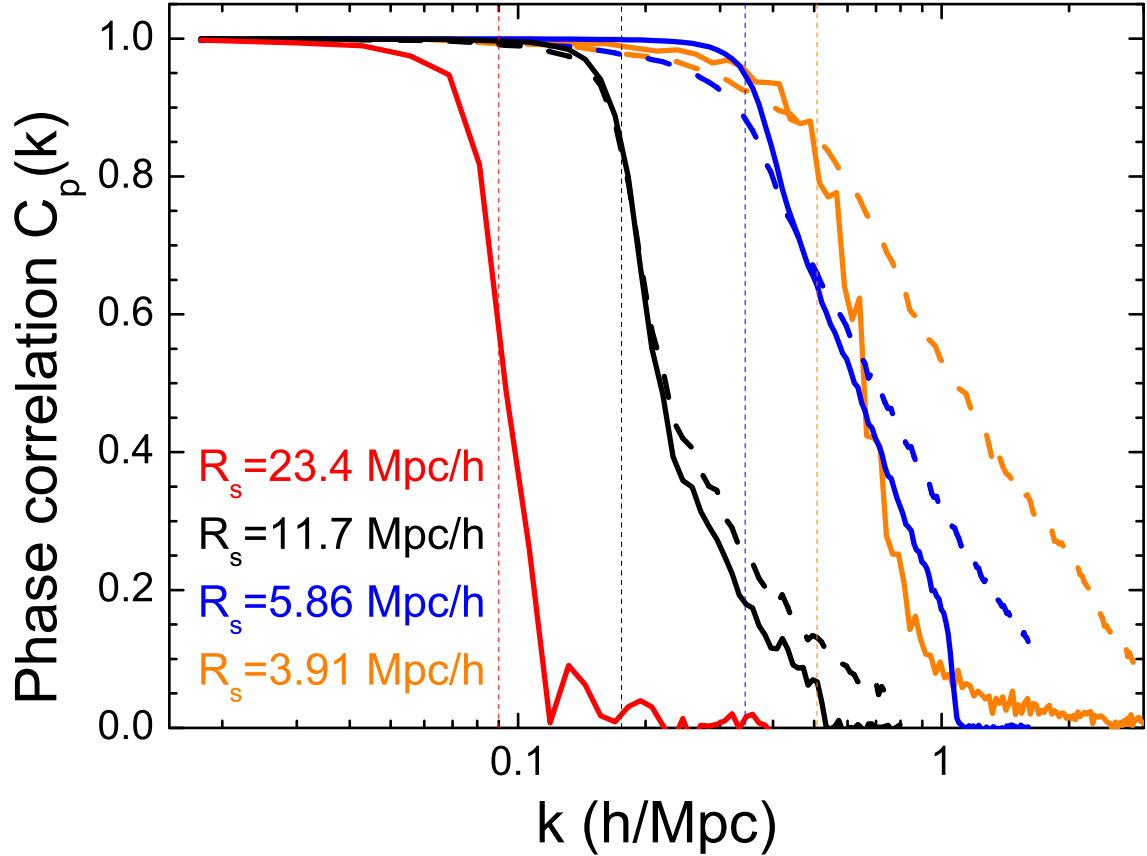,scale=1.} \caption{Solid lines: the phase
correlation between $\rho_{\rm mod}$ and $\rho_{\rm f}$ for tests
with different smoothing scales as indicated in the figure. Dashed
lines: the phase correlation between $\rho_{\rm rs}$ and
$\rho_{\rm f}$. $\rho_{\rm f}$ is the density of original
simulation, $\rho_{\rm mod}$ is the modeled density field and
$\rho_{\rm rs}$ is the resimulated density field. The vertical
dashed lines indicate the characteristic wavenumber $k_{\rm c}$
(see Section \ref{sec:sim} for the definition).} \label{fig_phcr}
\end{figure*}

\begin{figure*}
\epsfig{file=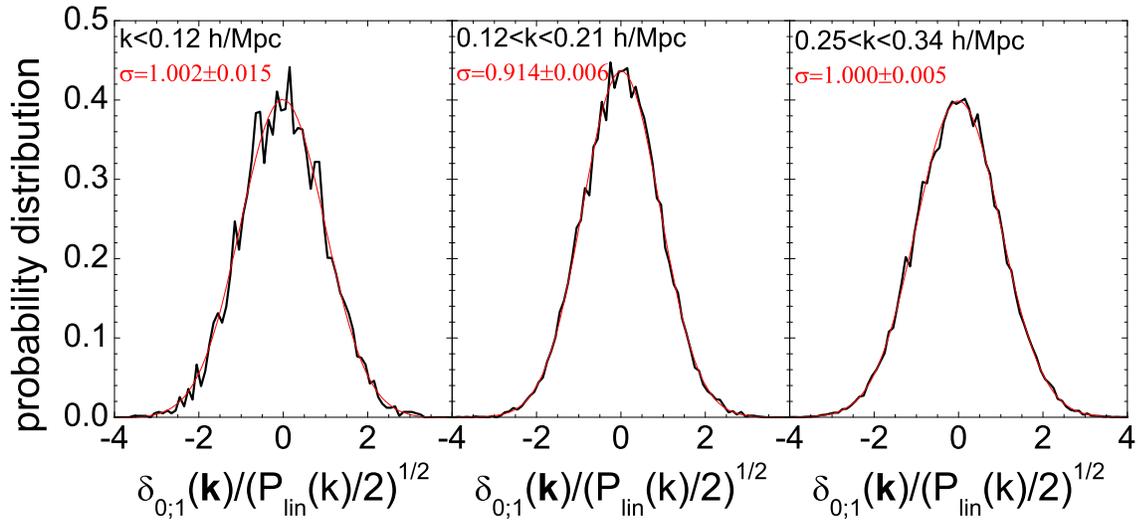,scale=1.} \caption{The black lines show
the probability distribution of $\delta_{0;1}({\bf
k})/\sqrt{P_{\rm lin}(k)/2}$ of the primary test at three different
scales. For comparison, we also show the best-fitting Gaussian
curves and their $\sigma$ in red.} \label{fig_gaudis}
\end{figure*}

\begin{figure*}
\epsfig{file=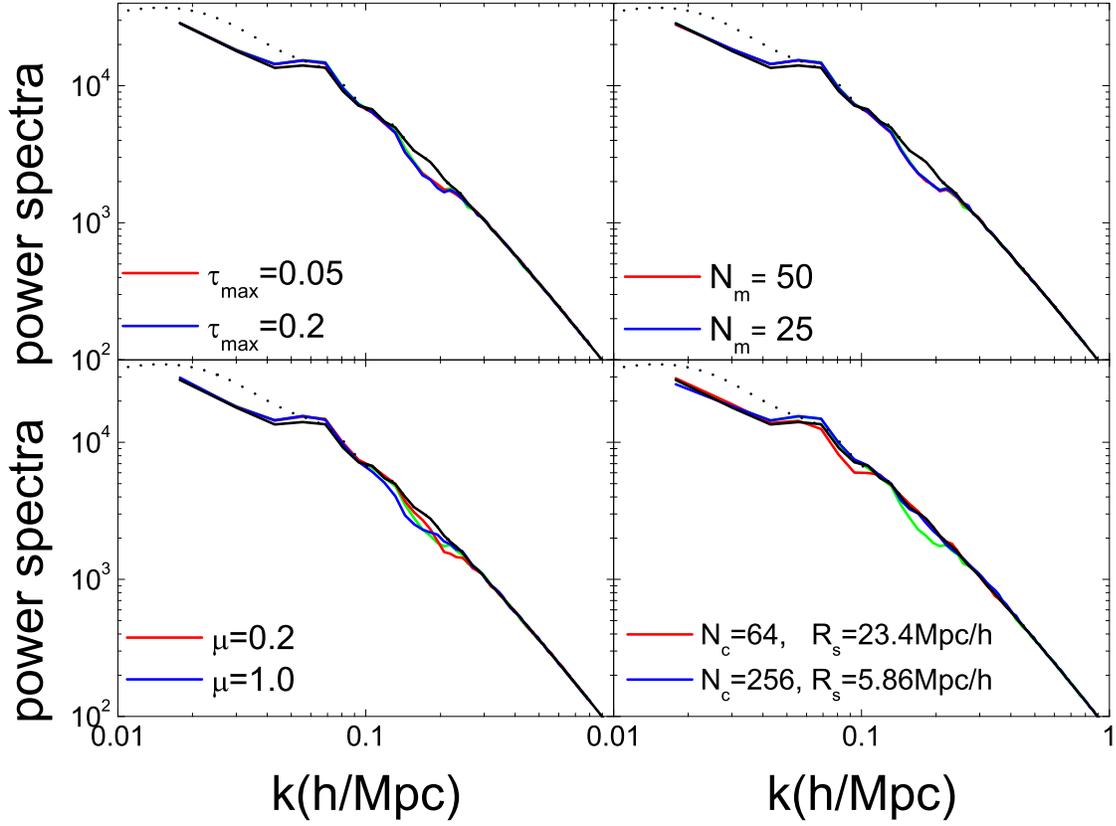,scale=1.} \caption{Power spectra for various
tests. In all four panels, dotted lines show the analytic linear
power spectrum, $P_{\rm lin}(k)$, the black solid lines show the
power spectrum measured from the MS simulation and the green lines
show the convergent power spectrum of primary test. The labels in
each panel show the parameters \emph{different} from the primary
test for the corresponding test.} \label{fig_dpps}
\end{figure*}

\begin{figure*}
\epsfig{file=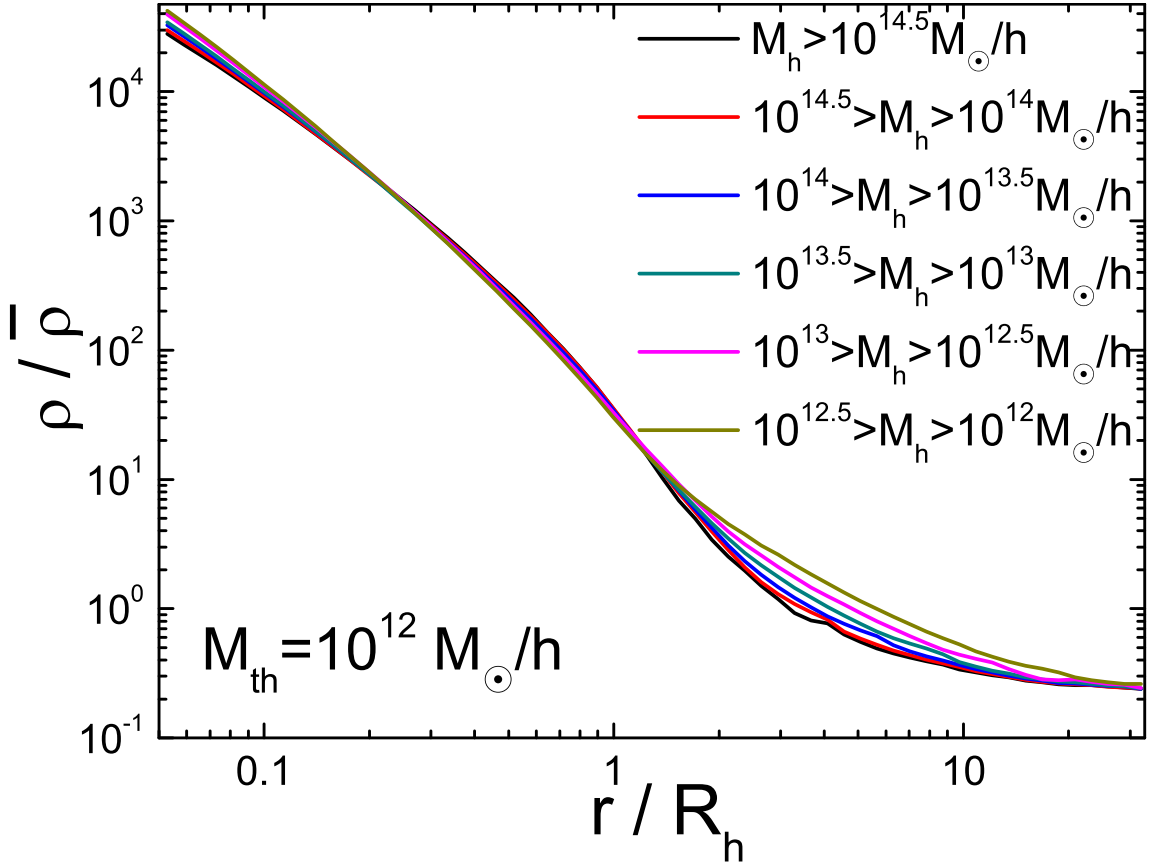,scale=1.} \caption{The density profiles of
mass in and around the haloes in various mass bins. Here the mass
threshold for the halo population is M$_{\rm th}$=$10^{12}\msun$.
The radius $r$ is scaled by halo virial radius $R_h$, and the
density is scaled with $\bar\rho$, the mean density of the
universe.} \label{fig_dp12}
\end{figure*}

\begin{figure*}
\epsfig{file=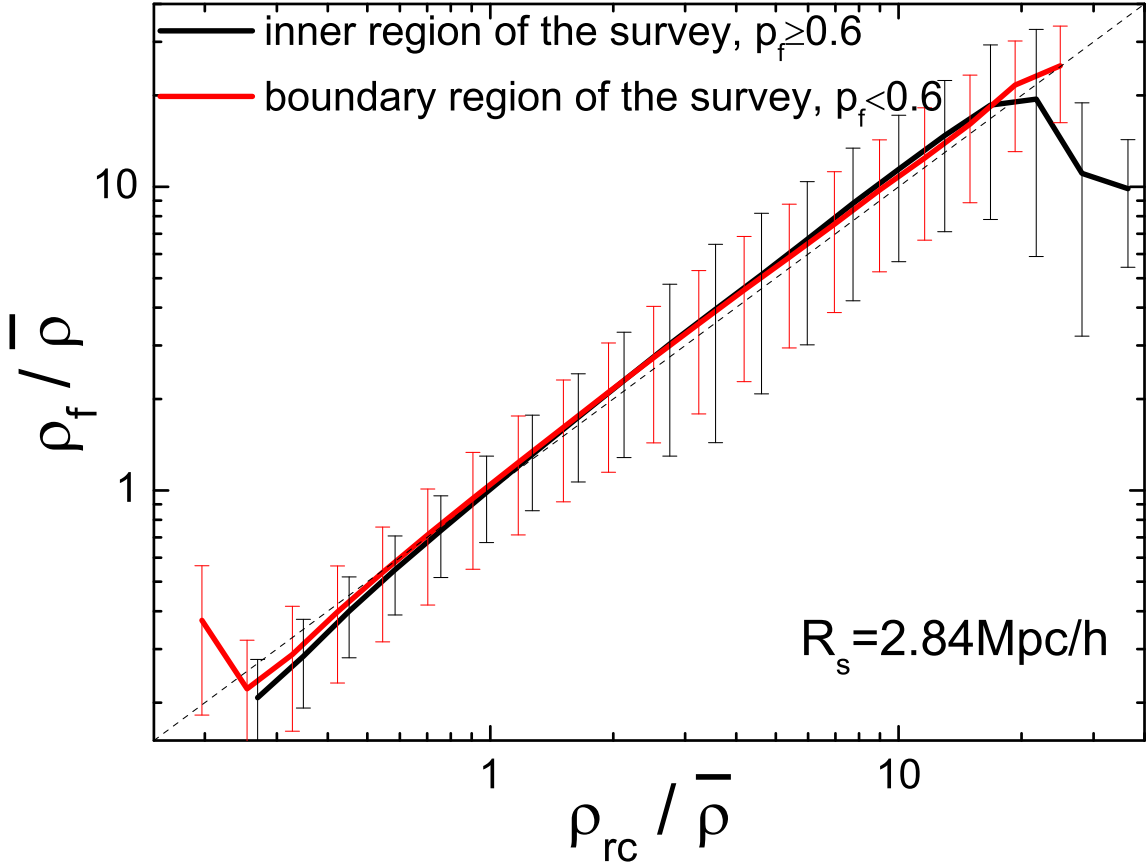,scale=1.} \caption{The comparison of
density between the original simulation and the reconstruction.
The reconstruction here is obtained by using mock groups with
M$_{\rm th}$=$10^{12}\msun$ and density profiles shown in Fig.
\ref{fig_dp12}. The two density fields are smoothed with Gaussian
kernel with smoothing scale of $2.84\mpc$. The black and red lines
show the results for the grids with different filling factor,
$p_{\rm f}$ (see text for definition).} \label{fig_rhorc}
\end{figure*}

\begin{figure*}
\epsfig{file=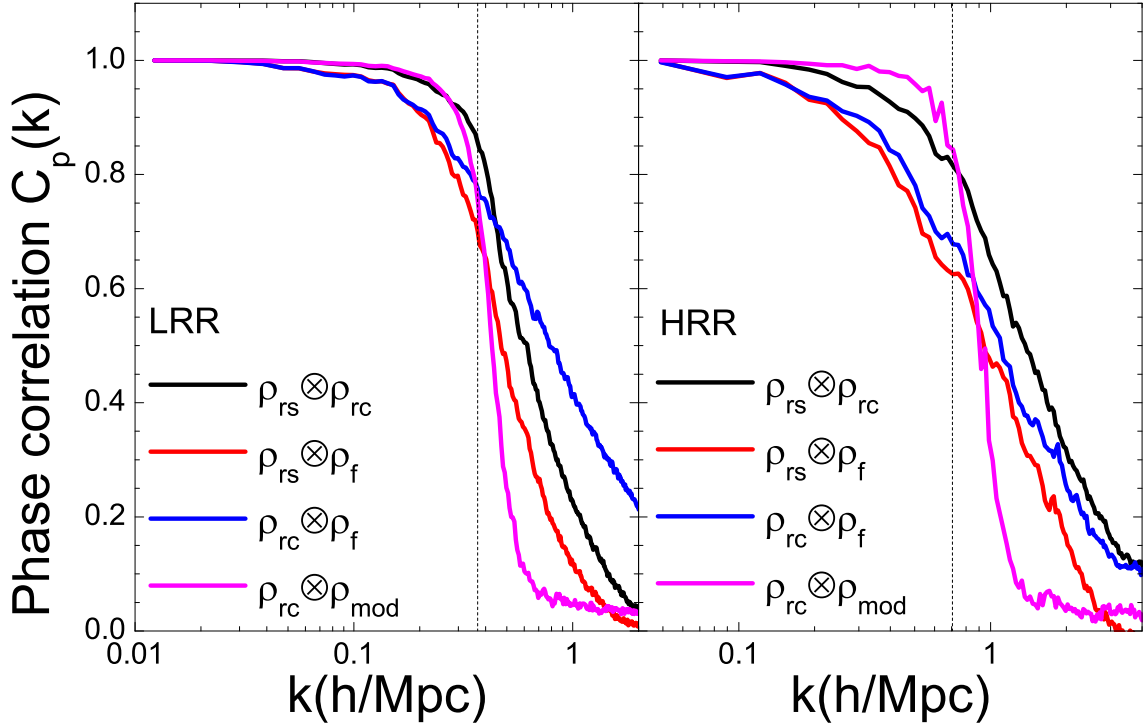,scale=1.} \caption{The phase correlation
among modeled density $\rho_{\rm mod}$, resimulated density
$\rho_{\rm rs}$, reconstructed density $\rho_{\rm rc}$ and
original density $\rho_{\rm f}$. The left panel shows the results
for LRR. The right panel shows the results for HRR. The vertical
dashed lines denote the characteristic wavenumber, $k_{\rm c}$.}
\label{fig_phrs}
\end{figure*}

\begin{figure*}
\epsfig{file=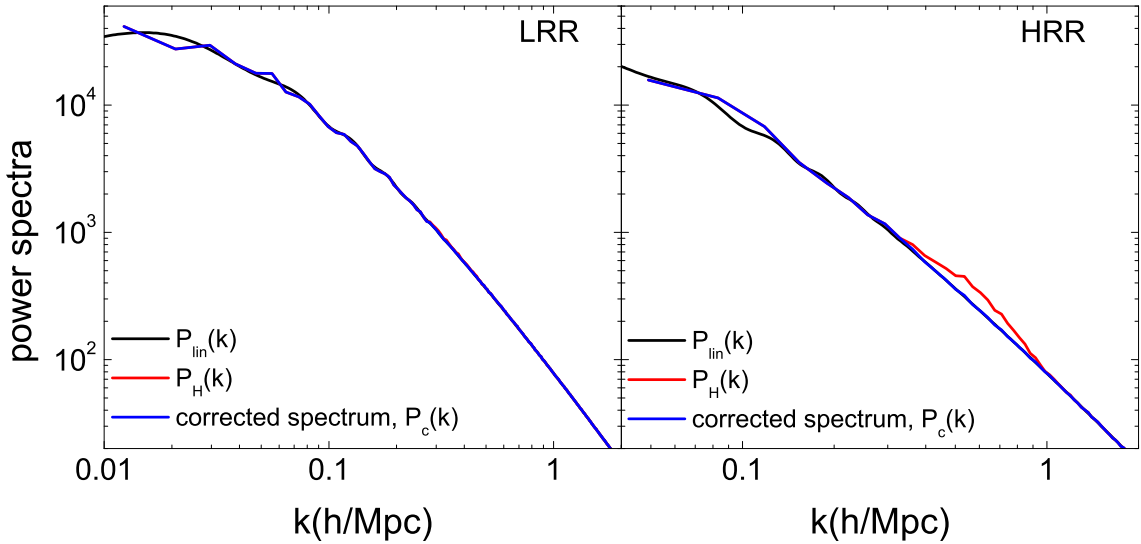,scale=1.} \caption{Black lines show the
analytic linear power spectrum, $P_{\rm lin}(k)$, the red lines show one
convergent Hamiltonian spectrum, $P_{\rm H}(k)$, and the blue lines show
corrected power spectrum, $P_{\rm c}(k)$. The left and right panels show
the results for LRR and HRR, respectively.} \label{fig_ps}
\end{figure*}

\begin{figure*}
\epsfig{file=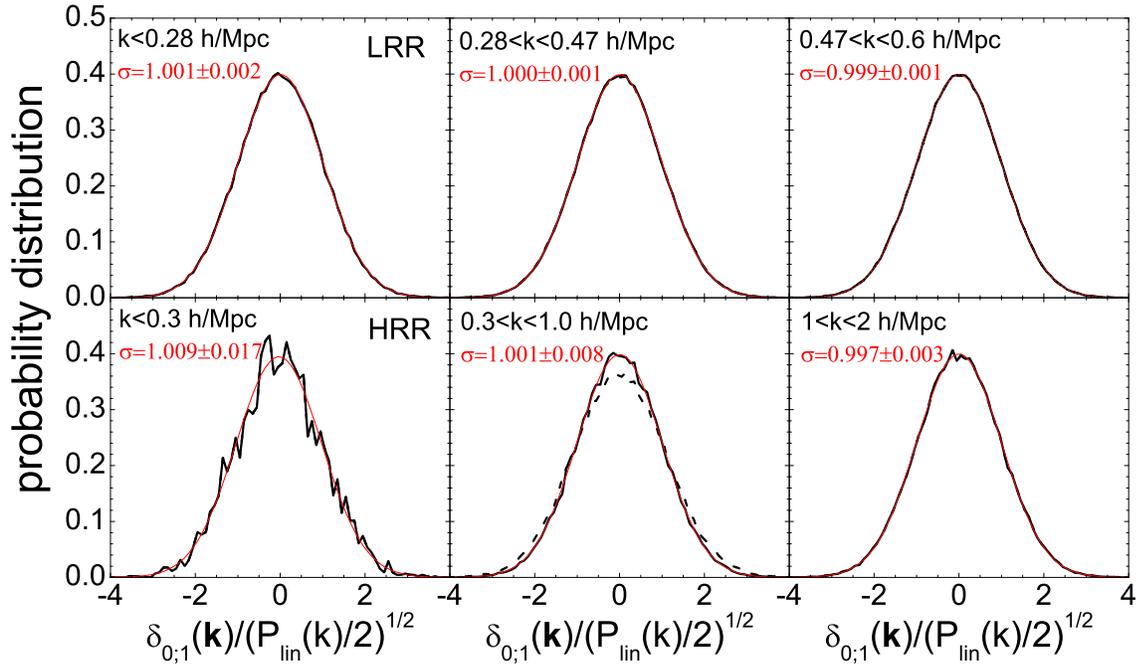,scale=1.} \caption{The probability
distribution of $\delta_{0;1}({\bf k})/\sqrt{P_{\rm lin}(k)/2}$ at large
(left), intermediate (middle) and small (right) scales for LRR
(upper) and HRR (lower). The dashed and solid black lines show the
results before and after correcting for the discrepancy,
respectively. The dashed and solid lines are undistinguishable at
all panels except the lower middle panel. The red lines are the
best-fitting Gaussian curves of the solid black lines. We also
present the best-fitting $\sigma$ in red digits.} \label{fig_mgau}
\end{figure*}

We only show the results at the 1700th chain steps for both the
simulated and reconstructed density fields in all figures (except
in Fig. \ref{fig_ins} and \ref{fig_evps} where results are shown
for different steps). Because we just want to reconstruct the
linear density field rather than to draw a sample for statistical
study (see Section \ref{sec_goal} for our objective in detail), it
is not necessary to show the results for all chain steps. In fact,
the accepted samples after burn-in have very similar Hamiltonian
power spectra and $\chi^2_{w}$ (see e.g. Figs. \ref{fig_ins} and
\ref{fig_evps}), the results at all other steps are very similar
to those that are shown. The choice of the 1700th step is
arbitrary; the only requirement is that the chain at this step has
already converged. Moreover, in order to investigate whether the
results are sensitive to the choice of the initial set of
$\delta_j({\bf k})$, we have performed tests with different
initial sets of $\delta_j({\bf k})$ and found the results change
very little. For example, the changes in $\chi^2_w$ and the phase
correlation functions are small.

\section{Re-simulations of Reconstructed Linear Density Field}
\label{sec:resim}

Up to now, all our results are based on the structure formation
model of TZ12. The advantage of using the TZ12 model is that it is
very fast and can thus be implemented into our HMC to infer the
initial linear density field. However, the TZ12 model is not
expected to work accurately in the highly non-linear regime.
Although the modeled density field shown above is closely
correlated with the input present-day density, it is unclear to
which extent the density field \emph{fully} evolved from our
reconstructed linear density field matches the original simulated
density field, especially on small scales where non-linear
evolution becomes important. In order to test the model in the
highly nonlinear regime, we need to use full $N$-body simulations
to evolve our reconstructed initial linear density field to the
present day and to compare the resimulation with the original
simulation.

To do this we set up the initial condition for our constrained N-body
simulation using the renormalized $\delta({\bf k})$. We generate the
initial displacement field at a given high redshift, $z_i$, using the
Zel'dovich approximation. The displacement field is used to perturb
the positions of the N-body particles that initially have uniform
distribution. Each particle is assigned a velocity according to the
growing mode solution of the linear perturbations. We then use the
N-body code Gadget-2 (Springel 2005) to evolve the initial condition
to the present day. The fully evolved density field is referred to as
the resimulated density field, and denoted by $\rho_{\rm rs}$.

\subsection{Initial Conditions from Simulated Density Field}

Let us first consider $\delta({\bf k})$ reconstructed from the
original simulated density field, i.e. $\rho_{\rm
p}\equiv\rho_{\rm f}$ (Section \ref{sec:sim}). For our primary test, we use $N_p=128^3$
particles in a box of $L=500\mpc$ to trace the evolution of the
density field. As shown in Table. \ref{tab_hmc}, we adopt an
initial redshift $z_i = 20$, a particle mass $m_p \simeq 4.14
\times 10^{12}\msun$ and a force softening length $\epsilon =
80\kpc$. The dashed black line in Fig. \ref{fig_phcr} shows the
phase correlation between $\rho_{\rm rs}$ and $\rho_{\rm f}$. As
can be seen, this phase correlation is very similar to that
between $\rho_{\rm mod}$ and $\rho_{\rm f}$. We also show the
results of the resimulation from the reconstructed linear density
field using $R_{\rm s} = 5.86\mpc$ and $3.91\mpc$, respectively.
For the test with $R_{\rm s}=5.86\mpc$, the phase correlation is
again very similar to that between $\rho_{\rm
  mod}$ and $\rho_{\rm f}$, suggesting that the MZA works well up to
$k\simgt 0.3\mpci$. In the test with $R_{\rm s}=3.91\mpc$,
however, the $\rho_{\rm rs}$ - $\rho_{\rm f}$ correlation is much
stronger than the $\rho_{\rm mod}$ - $\rho_{\rm f}$ correlation at
$k>k_{\rm c}$.

Given that the HMC method tends to minimize the difference between
$\rho_{\rm mod}$ and $\rho_{\rm f}$, it is unexpected that the
$\rho_{\rm mod}$ - $\rho_{\rm f}$ correlation is much weaker than
the $\rho_{\rm rs}$ - $\rho_{\rm f}$ correlation at $k>k_{\rm c}$.
The most important difference between $\rho_{\rm mod}$ and
$\rho_{\rm rs}$ is that the later is the result of fully
non-linear process, in which the mode of the non-linear density
field on small scales may be coupled to that on large scales (see,
e.g., Tassev \& Zaldarriaga 2012a). Consequently, part of phases
at $k>k_{\rm c}$, where the initial phases are not well
constrained, is reproduced in the re-simulation. Such mode
coupling is not included in the TZ12 model based on quasi-linear
theory, so that the phase information in $\rho_{\rm f}$ at $k\gg
k_{\rm c}$ is almost completely lost in $\rho_{\rm mod}$.

\begin{figure*}
\epsfig{file=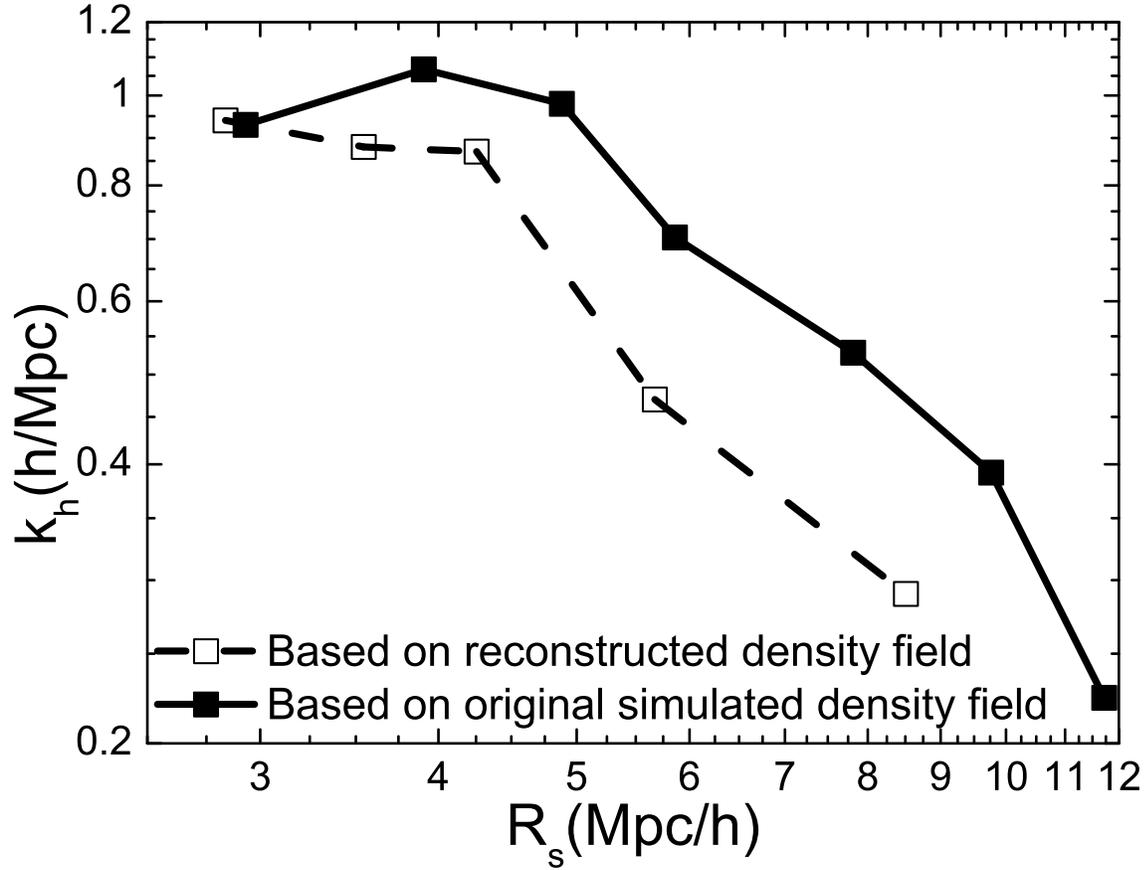,scale=1.} \caption{The figure show $k_{\rm
h}$, at which the phase correlation between the resimulated
density field and original simulated density field is half, as a
function of $R_{\rm s}$. The solid line show the results for the
re-simulations constrained directly from original simulation, while
the dashed line show the results for resimulations constrained
from the reconstructed density field.} \label{fig_rskh}
\end{figure*}

Our above results clearly show that the match between the
resimulation and the original simulation is better at smaller
$R_{\rm s}$. To quantify this trend, we introduce a quantity
$k_{\rm h}$ to measure the scale at which $\rho_{\rm rs}$ can well
match $\rho_{\rm f}$. It is defined in such a way that the phase
correlation between the two density fields at $k_{\rm h}$ is half.
In Fig. \ref{fig_rskh}, we show $k_{\rm h}$ as a function of
$R_{\rm s}$. On large scales, $k_{\rm h}$ increases with
decreasing $R_{\rm s}$, consistent with expectation because
information of the density field on scales below $R_{\rm s}$ is
lost due to the smoothing. As it reaches about unity, however, the
value of $k_{\rm h}$ becomes insensitive to $R_{\rm s}$. This is
also expected, because the TZ12 model is not expected to work
accurately on very small scales. This demonstrates that in order
to reconstruct the density field on scales below $k\sim 1\mpci$,
one has to use a model that is more accurate than the TZ12 model
adopted here.

\subsection{Initial Conditions from Reconstructed Density Field}

\begin{figure*}
\epsfig{file=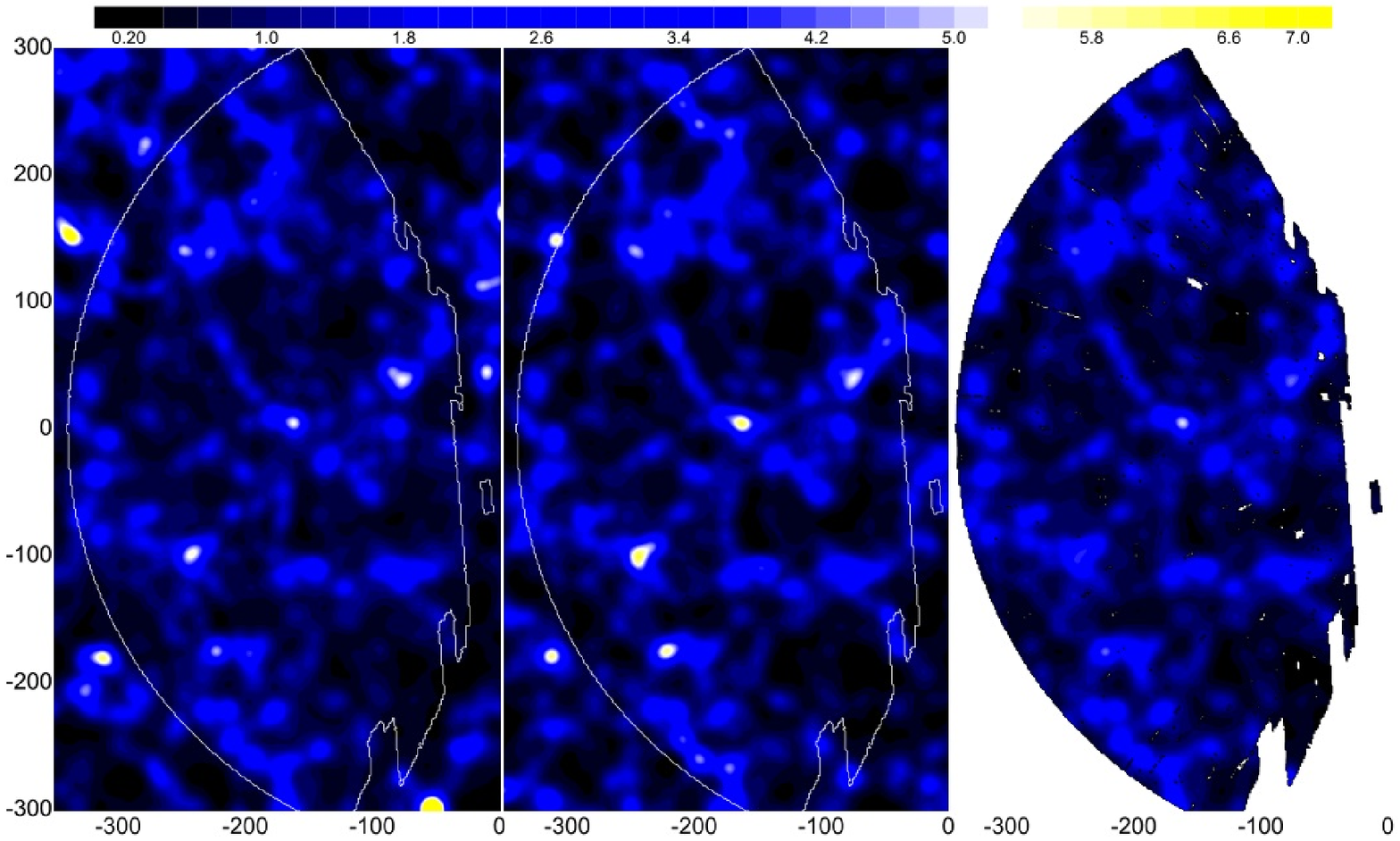,scale=0.8} \caption{The low-resolution density
maps in a slice of $600\times350\times4.25\mpc$. The left panel
shows the resimulated density map from LRR. The middle and right
panels show the original simulated and reconstructed density
fields. All these density fields are smoothed with Gaussian kernel
with $R_{\rm s}=5.67\mpc$ and scaled with the mean density of the
universe.} \label{fig_dcl}
\end{figure*}

\begin{figure*}
\epsfig{file=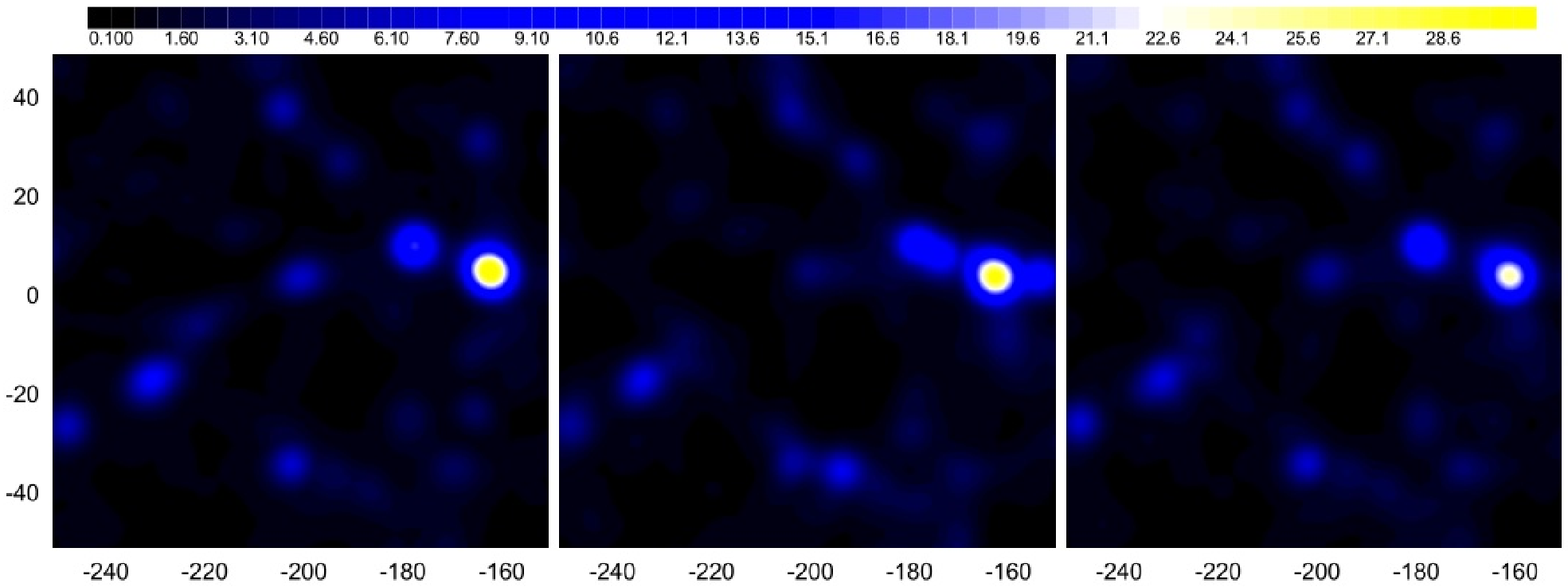,scale=0.8} \caption{The high-resolution
density maps in a slice of $100\times100\times6.4\mpc$. The left
panel shows the resimulated density field from HRR. The middle and
right panels show the original simulated and reconstructed density
fields. All these density fields are smoothed with Gaussian kernel
with $R_{\rm s}=2.84\mpc$ and scaled with the mean density of the
universe.} \label{fig_dcs}
\end{figure*}

In this subsection, we use $N$-body simulations to evolve the
initial conditions obtained from the present-day density
field reconstructed from the mock catalogs ($\rho_{\rm
p}\equiv\rho_{\rm rc}$, Section \ref{sec:mocktest}). For
reference we list the parameters for both LRR and HRR in Table
\ref{tab_hmc}. To inspect our results visually, we present density
maps of the same thin slices in the original simulation used to
construct the mock catalog, the density field reconstructed from
the mock catalog, and the density field in the re-simulations.
Fig. \ref{fig_dcl} shows the results for the LRR. Here all density
fields are smoothed with a Gaussian kernel with $R_{\rm
s}=5.67\mpc$. Within the survey volume, the three maps look quite
similar; almost all structures in the original simulation, such as
massive clusters, filaments and underdense voids, are reproduced
in the re-simulation. The density maps for the HRR, smoothed with
$R_{\rm s}=2.84\mpc$, are presented in Fig. \ref{fig_dcs}. There
are about twenty bright spots in the original simulation, which
correspond to a single halo or a cluster of a few haloes with
masses down to about a few times $10^{12}\msun$. Most of these
structures are clearly reproduced in our resimulation. In
addition, some small filaments in the original simulation are also
correctly reproduced in the re-simulation. This is quite
remarkable, given that our reconstruction from the mock catalog
uses only groups with assigned masses above $M_{\rm
th}=10^{12}\msun$ and that non-linear effects are important on
such small scales.

\begin{figure*}
\epsfig{file=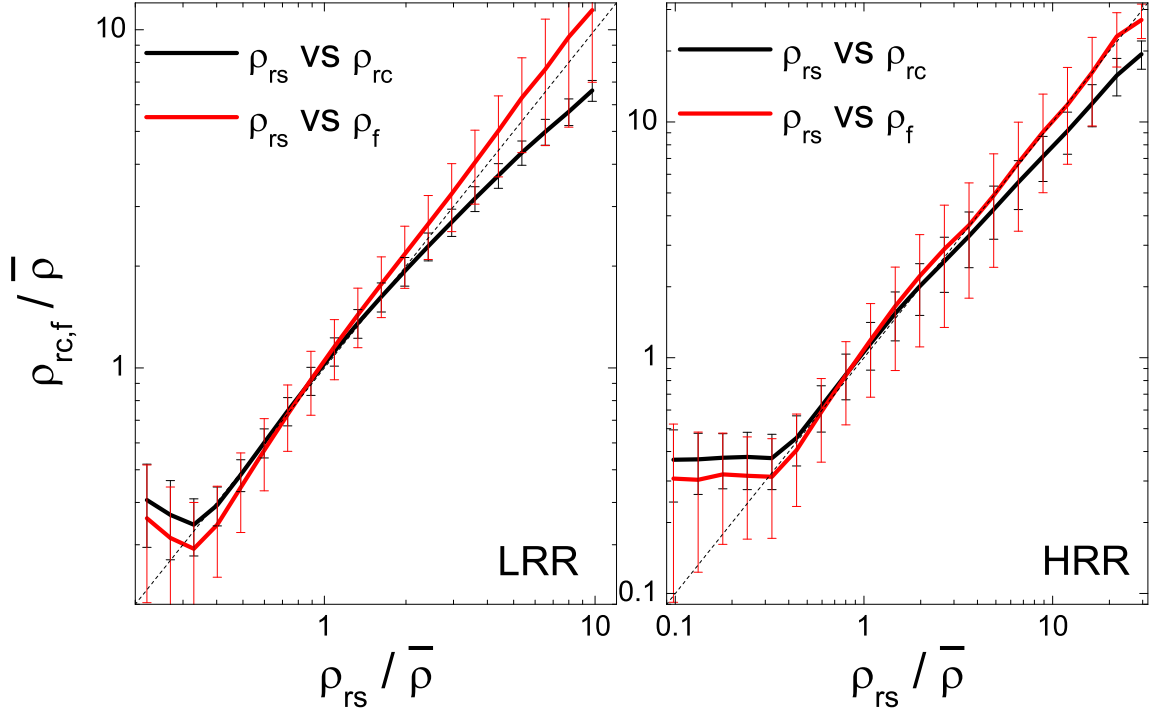,scale=1.} \caption{Black lines show the
comparison between $\rho_{\rm rs}$ and $\rho_{\rm rc}$. Red lines
show the comparison between $\rho_{\rm rs}$ and $\rho_{\rm f}$.
All these densities are scaled with $\bar\rho$, the mean density
of the universe. The left panel shows the results for LRR and the
densities are smoothed with Gaussian kernel with $R_{\rm
s}=5.67\mpc$. The right panel shows the results for HRR and the
densities are smoothed with Gaussian kernel with $R_{\rm
s}=2.84\mpc$. The dot lines indicate the one to one
relation.}\label{fig_rhors}
\end{figure*}

In the left panel of Fig. \ref{fig_rhors} we show the comparison of
$\rho_{\rm rs}$ with $\rho_{\rm rc}$ and $\rho_{\rm f}$ for the
LRR. There is weak bias at the high density end in both relations:
while $\rho_{\rm rs}$ is higher than $\rho_{\rm rc}$, it is slightly
lower than $\rho_{\rm f}$. Since the initial condition for the
resimulation is constrained by the reconstructed density field, the
$\rho_{\rm rs}$ - $\rho_{\rm rc}$ relation is much tighter than the
$\rho_{\rm rs}$ - $\rho_{\rm f}$ relation. The typical dispersion in
the former relation is about 0.05 dex, while it is about two to three
times larger in the latter relation. The phase correlations among the
three density fields are shown in the left panel of Fig.
\ref{fig_phrs}. Similar to the $\rho_{\rm mod}$ - $\rho_{\rm rc}$
correlation, there is a sharp transition in the correlation
coefficient from unity to about zero around $k_{\rm c}$. Upon closer
examination, we find that the $\rho_{\rm rs}$ - $\rho_{\rm f}$ phase
correlation is always lower than the minimum of the $\rho_{\rm rs}$ -
$\rho_{\rm rc}$ and $\rho_{\rm rc}$ - $\rho_{\rm f}$ phase
correlations. This is expected, as the accuracy of the resimulation
depends both on the accuracy of the HMC method, which determines the
strength of the $\rho_{\rm rs}$-$\rho_{\rm rc}$ correlation, and the
accuracy in the reconstruction of the present-day density field, which
determines the strength of the $\rho_{\rm rc}$-$\rho_{\rm f}$
correlation.

The comparisons of $\rho_{\rm rs}$ with $\rho_{\rm rc}$ and $\rho_{\rm f}$
for the HRR are shown in the right panel of Fig. \ref{fig_rhors}. As
one can see, $\rho_{\rm rs}$ is also strongly correlated with both
$\rho_{\rm rc}$ and $\rho_{\rm f}$. The scatter in the $\rho_{\rm
  rs}$-$\rho_{\rm f}$ correlation is less than 0.2 dex in most density
bins. At the high-density end, a weak bias is present in the
$\rho_{\rm rs}$- $\rho_{\rm rc}$ relation, but such a bias is
absent in the $\rho_{\rm rs}$ - $\rho_{\rm f}$ relation. At the
low-density end ($\rho <0.3\bar\rho$), $\rho_{\rm rs}$ is
correlated neither with $\rho_{\rm rc}$ nor $\rho_{\rm f}$. The
reason is that in our reconstruction of the present-day density
field using mass distributions around haloes, the minimum of the
density profiles in the domain is about $0.25\bar\rho$ (Fig.
\ref{fig_dp12}).  Therefore the information in the most underdense
regions is totally lost when reconstructing the present-day
density field.  This bias can be mitigated by adopting a smaller
mass threshold $M_{\rm th}$ for the group catalogue. The usage of
a smaller $M_{\rm th}$ can lower the minimum density that our
reconstruction can reach (see Wang et al. 2009a).

The phase correlations for the HRR are presented in the right
panel of Fig. \ref{fig_phrs}. The phase correlation between
$\rho_{\rm rs}$ and $\rho_{\rm rc}$ declines gradually around
$k_{\rm c}$.  Similar to the results shown in the previous
subsection, this correlation is much stronger than the $\rho_{\rm
mod}$ - $\rho_{\rm rc}$ correlation at $k>k_{\rm c}$. As discussed
above, this is due to mode coupling which is not fully included in
the TZ12 model. Moreover, the $\rho_{\rm rc}$ - $\rho_{\rm f}$
phase correlation lies below the $\rho_{\rm rs}$ - $\rho_{\rm rc}$
phase correlation at almost all scales.  It indicates that, on the
smoothing scale $R_{\rm s}=2.84\mpc$, the accuracy of the
re-simulation in matching the original density field is mainly
limited by the reconstruction of the present day density rather
than by the HMC method. Despite all of these, the match between
the resimulation and the original simulation is remarkable. At
$k=k_{\rm c}$, the phase correlation between $\rho_{\rm rs}$ and
$\rho_{\rm f}$ is still as high as 0.6; even at $k=1.0\mpci$,
about half ($47\%$) of the phase information is reproduced. We
recall again that the translinear scale corresponds to
$k=0.15\mpci$.

We also perform tests with other values of $R_{\rm s}$ and measure
$k_{\rm h}$, at which the phase correlation between the
resimulation and original simulation is half. We show $k_{\rm h}$
as a function of $R_{\rm s}$ as the dashed line in Fig.
\ref{fig_rskh}. One see that the value of $k_{\rm h}$ first
increases with decreasing $R_{\rm s}$ then remains almost
constant, and has the maximum of $0.94\mpci$ at $R_{\rm
s}=2.84\mpc$. The curve generally lies below that based on
original simulated density field (solid line), because the
reconstruction of the present-day density field is not perfect.
Note that at $R_{\rm s}=2.84\mpc$, the value of $k_{\rm h}$
obtained from the reconstructed present-day density field is
similar to that obtained from the simulated present-day density
field, because on such small scales the inaccuracy of the TZ12
model becomes the dominating source of error in the reconstructed
initial condition.

\section{Summary and Discussion}
\label{sec:discussion}

In this paper we have developed an
effective method to reconstruct the
linear density field that underlies the formation of the cosmic
density field in the local Universe. To this end we have developed
a Hamiltonian Markov Chain Monte Carlo (HMC) method which allows
us to generate the linear density field based on a
posterior probability function. This distribution consists of two
components, a prior term which takes into account the Gaussianity
and power spectrum assumed for the linear density field, and a
likelihood term designed to ensure that the predicted density
field from the initial condition matches a given final density
field. We adopt the modified Zel'dovich approximation developed by
TZ12 to model the final, evolved density field off the initial,
linear density field.

The HMC method is based on an analogy to a
physical system. The potential is taken to be minus the logarithm
of the posterior function. The momenta are drawn from given
Gaussian distributions before each chain step so that the
fictitious system can continue to equilibrate and ``orbit" within
the extended potential well with the passage of ``time". The
system eventually converges to a state in which balance between
kinetic and potential ``energy" is achieved. Using a simulated
density field, we demonstrate that our HMC method converges very
quickly, and that the converged linear density fields closely
follow a Gaussian distribution with a spectrum that accurately
matches the input linear power spectrum. A small discrepancy is
found at the scales where the likelihood and prior terms in the
Hamiltonian force are comparable. This discrepancy, however, can
straightforwardly be corrected for by re-normalizing the
amplitudes of the corresponding Fourier modes (while keeping their
phases fixed) with the input linear spectrum. We find that the
modeled density field matches the input density field with high
precision, with a RMS difference typically smaller than $5\%$.

Since our HMC method needs the present-day density field as a
constraint in reconstructing the initial linear density field, we
also present a method to reconstruct the present-day density field
from mock galaxy and group catalogues. The mock catalogues are
constructed from the MS simulation for the SDSS DR7, taking
detailed account of the angular variation of the magnitude limit
and the survey completeness (Wang et al. 2012a and the references
therein) so that we can verify the reliability of our method in
real applications. We use the method developed by Wang et al.
(2009a) to reconstruct the density field based on the mock group
catalogue, taking into account inaccuracies in the group finder,
as well as uncertainties arising from the assignment of a halo
mass to each individual group and the redshift space distortions.
We find that the phase correlation between the reconstructed and
simulated density fields is almost perfect at large scales, with a
correlation coefficient close to one. The scale at which the
correlation coefficient drops to 0.5 is $k \sim 1\mpci$,
indicating that our method works surprisingly well down to scales
that are well into the non-linear regime.

We apply the HMC method to the reconstructed density field in two
different volumes. The first one (LRR) is the entire survey volume
of the SDSS embedded in a periodic box of size $726\mpc$, and the
second one (HRR) is a cubic box of size $100\mpc$ covering the
inner region of the survey volume. These two applications are used
to test and verify the performance of the HMC method over a wide
dynamic range. As an additional test of the performance of our
methods, we use the reconstructed linear density fields of LRR and
HRR to generate initial conditions, which we subsequently evolve
to the present day using a $N$-body simulation code. Both visual
inspection and quantitative analysis show that the density field
obtained from these re-simulations accurately match the density
field of the original simulation used to construct the mock
catalog. In particular, the phase correlation between
re-simulation and original simulation has a coefficient close to
unity on large scales and only starts to drop to 0.5 at $k \sim
1\mpci$. This clearly demonstrates that our HMC method together
with the reconstruction method of Wang et al. (2009a) provides a
robust way to reconstruct the initial conditions for the local
cosmological density field from observational data.

Numerous studies in the past have tried to infer the initial
conditions of structure formation in the local Universe using
observational data such as galaxy distributions and/or peculiar
velocity surveys (e.g. Nusser \& Dekel 1992; Weinberg 1992; Kolatt
et al. 1996; Klypin et al.  2003; Dolag et al. 2005; Lavaux 2010).
Most of these studies integrate the observed density field
\emph{backwards} in time to some initial time. However, these
approaches suffer from complications in the observational data,
such as spatial variations in the magnitude limit and complex
survey boundaries, as well as the amplification of noise and
numerical errors by the decaying mode during backward integration
(Nusser \& Dekel 1992). As pointed out by JW13, these problems can
be overcome by the HMC method. The amplification of noise
and numerical errors is not an issue since the HMC method uses
\emph{forward} evolution of the cosmic density field (see also
Kitaura 2013), and the survey geometry is taken into account by
the weight field in the likelihood. Finally, it is
worth emphasizing that some of the previous methods had to
Gaussianize the inferred initial density field using some
order-preserving transformation. In the HMC method adopted here,
however, the initial density field is Gaussian by the construction
of the posterior.

Our own HMC method has some unique advantages. We design the
likelihood using the present-day density field reconstructed from
galaxy groups (i.e., dark matter haloes), rather than the galaxy
distribution itself. The latter requires a detailed understanding
of how galaxies are biased relative to the underlying dark matter
distribution (see e.g. Kitaura \& En{\ss}lin 2008). As mentioned
earlier, this bias between galaxies and dark matter is far from
trivial; it depends on galaxy properties, has stochastic and
non-linear contributions, and may even be non-local. Adopting a
simple linear bias model would significantly underestimate the
density in high density regions. Currently it is still unclear how
to directly use the galaxy distribution to model the density
field, especially in high density region, in an unbiased way.
Another problem with using the galaxy distribution is that the
constraint becomes very poor in underdense regions, where only few
or no galaxies can be observed in a uniformly selected galaxy
sample (JW13). In our approach, these problems are largely absent.
As shown in Section \ref{sec_denrc}, the reconstructed density
fields based on the detailed mock catalogues match the input
density field very well over a large dynamical range. Furthermore,
since our reconstruction relies on groups (haloes), we can
accurately reproduce the high density regions within individual
haloes. In underdense regions, using the density profiles in the
domains of haloes can recover the density field down to $\simlt
0.25 \bar\rho$. And in principle, we can recover even lower
densities by simply using groups (haloes) above a lower mass
threshold, although this requires either a deeper redshift survey,
or the use of a more limited volume. The
use of groups to trace the large-scale density field
can effectively mitigate the problem due to the
Finger-of-God effect, which may severely impact
the reconstruction if not properly handled. The reliability of our
reconstruction is further demonstrated by the fact that the
resimulations from the initial conditions constrained by the
reconstruction match the original simulation remarkably well in
the same density range.

Moreover, our HMC method works in Fourier space. Different Fourier
modes are mutually independent in the prior, and so are the
real and imaginary parts of individual modes. This enables us to
derive simple formulae for both the prior and likelihood terms in
the Hamiltonian force, and makes the computation much faster. As
shown in Table \ref{tab_hmc}, it takes, on average, only about 21,
220 and 2100 seconds for each step for $N_{\rm c}=128$, $256$ and
$512$ respectively. Particularly, our method can successfully
handle more than one hundred million free parameters! Such
efficiency is crucial when aiming to achieve high resolution in a
large volume.

In forthcoming papers, we will apply our reconstruction and HMC
methods to the SDSS DR7 group catalogue in order to generate the
initial conditions for structure formation in the SDSS volume.  We
will then use these initial conditions to run constrained
simulations to study the evolution of the local cosmic density
field.  This will provide a unique opportunity to further our
understanding of the formation and evolution of the galaxies we
directly observe.  For example, one can investigate the
correlation between the large scale environments, measured from
the constrained simulation, and the observational galaxy
properties. Recent studies have found significant dependencies of
halo properties on the large-scale environments, in particular the
large scale tidal fields (see e.g. Wang et al. 2011 and the
references therein), and it would be interesting to see whether
this is also the case for galaxies that reside in haloes. One can
also perform semi-analytical models of galaxy formation using halo
merger trees extracted directly from the constrained simulations.
The comparison between model galaxies and real galaxies in the
same large scale structures, such as filaments, sheets and
clusters, will provide us an avenue to constrain galaxy formation
in a way that is free of cosmic variance.

Finally, the constrained simulation can also be used to study the
physics and dynamics of the IGM. For instance, the hot gas and
peculiar velocities predicted by the constrained simulations can
be used to make predictions for the (both thermal and kinetic)
Sunyaev-Zel'dovich effects, which can be compared with
(forthcoming) observations. Moreover, a comparison of the
simulated density field with quasar absorption lines in a wide
range of ionization potentials can provide constraints on the
metallicity and temperature of the baryonic gas inside cosmic web
(Cen \& Ostriker 1999). Such studies will provide a unique way for
understanding the nature of the low-$z$ absorption systems and the
state and structure of the IGM. In particular, it will allow a
detailed exploration of the connection and interaction between the
galaxy population and the IGM.

\section*{Acknowledgments}

This work is supported by NSFC (11073017, 10925314, 11128306,
11121062, 11233005), NCET-11-0879 and the Fundamental Research
Funds for the Central Universities. HJM would like to acknowledge
the support of NSF AST-0908334 and the CAS/SAFEA International
Partnership  Program for Creative  Research Teams (KJCX2-YW-T23).

\newpage

\end{document}